\pdfoutput=1

\documentclass[ 
reprint,
superscriptaddress,
amsmath,amssymb,
aps,
prr,
longbibliography,
]{revtex4-1}

\usepackage{mathtools}
\usepackage{dsfont}                             
\usepackage{bm}                                 
\usepackage[makeroom]{cancel}                   

\usepackage{color}
\definecolor{tangoblue3}{rgb}{0.13,0.3,0.53}

\usepackage{hyperref}                          
\hypersetup{
     colorlinks   = true,
     citecolor    = tangoblue3,
     linkcolor    = tangoblue3}
\usepackage[nameinlink,capitalize]{cleveref}    

\usepackage{graphicx}
\DeclareGraphicsExtensions{.pdf, .png, .jpg, .eps}     

\usepackage{enumitem} 
    
\begin{document}

\title{Dynamics of random recurrent networks with correlated low-rank structure}
\author{Friedrich Schuessler}
\affiliation{Rappaport Faculty of Medicine and Network Biology Research Group,
    Technion -- Israel Institute of Technology, Haifa, Israel}
\author{Alexis Dubreuil}
\affiliation{%
Laboratoire de Neurosciences Cognitives et Computationnelles, INSERM U960, 
Ecole Normale Superieure - PSL Research University, 75005 Paris, France}
\author{Francesca Mastrogiuseppe}
\affiliation{Gatsby Computational Neuroscience Unit, UCL, London, Great Britain}
\author{Srdjan Ostojic}
\affiliation{%
Laboratoire de Neurosciences Cognitives et Computationnelles, INSERM U960, 
Ecole Normale Superieure - PSL Research University, 75005 Paris, France}
\author{Omri Barak}
\email[email: ]{omri.barak@gmail.com}
\affiliation{Rappaport Faculty of Medicine and Network Biology Research Group,
    Technion -- Israel Institute of Technology, Haifa, Israel}
\date{\today}

\begin{abstract}
A given neural network in the brain is involved in many different tasks. 
This implies that, when considering a specific task, the network's connectivity contains a component 
which is related to the task and another component which can be considered random. 
Understanding the interplay between the structured and random components, and their effect on network 
dynamics and functionality is an important open question. 
Recent studies addressed the co-existence of random and structured connectivity, 
but considered the two parts to be uncorrelated. 
This constraint limits the dynamics and leaves the random connectivity non-functional. 
Algorithms that train networks to perform specific tasks typically generate correlations
between structure and random connectivity. 
Here we study nonlinear networks with correlated structured and random
components, assuming the structure to have a low rank. We develop an analytic framework to establish the precise effect of the correlations on the eigenvalue
spectrum of the joint connectivity. We find that the spectrum consists of a bulk
and multiple outliers, whose location is predicted by our theory. Using mean-field theory, we show that these outliers directly determine
both the fixed points of the system and their stability. Taken together, our analysis elucidates how
correlations allow structured and random connectivity to synergistically extend the range of computations available to networks.
\end{abstract}

\pacs{Valid PACS appear here}
\maketitle 

\section{Introduction}%
\label{sec:introduction}
One of the central paradigms of neuroscience is that computational function determines
connectivity structure: if a neural network is involved in a given task, its connectivity
must be related to this task. However, a given circuit's connectivity 
also depends on development and the learning of a multitude of 
tasks \cite{rigotti2013importance, yang2019task}. 
Accordingly, connectivity has often been depicted as containing a sum of random and structured 
components \cite{rivkind2017local,mastrogiuseppe2018linking,tirozzi1991chaos,roudi2007balanced,%
ahmadian2015properties}. 
Given that structure emerges through adaptive processes on top of 
existing random connectivity, one would intuitively expect correlations between the two
components. Nevertheless, the functional effects of the interplay between the random and the 
structured components have not been fully elucidated.

Networks designed to solve specific tasks often use purely structured 
connectivity \cite{ben1995theory,hopfield1982neural,wang2002probabilistic} that has been analytically 
dissected \cite{amit1985spin}.
The dynamics of networks with purely random connectivity were also thoroughly explored, 
charting the transitions between chaotic and ordered activity 
regimes \cite{sompolinsky1988chaos,rajan2010stimulus,brunel2000dynamics,wainrib2013topological,%
van1996chaos,huang2019circuit}.
Adding \textit{uncorrelated} random connectivity to a structured one was shown to generate the 
activity statistics originating from the random component while retaining the functional 
aspects of the structured one \cite{mastrogiuseppe2018linking,tirozzi1991chaos,roudi2007balanced,%
renart2007mean}. 

A specific setting in which correlations between random and structured components arise is the 
training of initially random networks to perform tasks. One class of training algorithms, 
reservoir computing, only modifies a feedback loop on top of the initial random connectivity 
\cite{maass2002real,jaeger2004harnessing,sussillo2009generating}. 
These algorithms can be used to obtain a wide range of computations 
\cite{enel2016reservoir,barak2013fixed}.
Recently, a specific instance of a network trained to exhibit multiple fixed points was analytically examined \cite{rivkind2017local}. It was shown that 
the dependence between the feedback loop and the initial connectivity is essential to obtain the desired functionality, but the explicit form of the correlations and the manner in which they determine functionality remained elusive. 

Thus there is no general theory linking the correlations between random and structured components to network dynamics.
Here we address this issue by examining the nonlinear dynamics of networks with such correlations. Because the dynamics of nonlinear systems vary between different areas of phase space, we focus on linearized dynamics around different fixed points. To facilitate the analysis, we consider low-rank structured components
which were shown to allow for a wide range of functionalities \cite{mastrogiuseppe2018linking}.

We develop a mean field theory that takes into account correlations between the
random connectivity and the low-rank part. Our theory directly links these correlations to the spectrum of the
connectivity matrix. 
We show how a correlated rank-one perturbation can lead to multiple spectral outliers and fixed points, a phenomenon that requires high-rank perturbations in the uncorrelated case \cite{mastrogiuseppe2018linking}. We analytically study dynamics around non-trivial fixed points, revealing a surprising connection between the spectrum, the fixed 
points and their stability. 
Taken together, we show how correlations between the low-rank structure and the 
random connectivity extend the computations of the joint network beyond the sum 
of its parts. 

\section{Network model}%
\label{sec:network_model}
We examine the dynamics of recurrent neural networks with correlated random and structured components in their connectivity. The structured component $P$ is a low-rank matrix and the random component $J$ 
is a full-rank matrix.
Network dynamics with such a connectivity structure have been analyzed 
for $P$ being independent of the random connectivity \cite{mastrogiuseppe2018linking}.
The learning frameworks of 
echo state networks and FORCE 
also have such connectivity structure
\cite{jaeger2004harnessing,sussillo2009generating}. 
There, however, the structure $P$ is trained
such that the full network performs a desired computation, 
possibly correlating $P$ to $J$. 

For most of this study, we set the rank of $P$ to one and write it as the outer
product 
\begin{equation}
    P = \mathbf{mn}^T
\end{equation}
of the two structure vectors $\mathbf{m}$ and $\mathbf{n}$. 
The matrix $J$ and vector $\mathbf{m}$ are drawn independently from normal distributions, 
$J_{ij} \sim \mathcal{N}(0, g^2 / N)$ and $m_i \sim \mathcal{N}(0, 1)$, where $N$ is the network size and $g$ controls the strength of the random part
\cite{sompolinsky1988chaos}. The second vector $\mathbf{n}$ is 
defined in terms of $J$ and $\mathbf{m}$. In this sense, $\mathbf{n}$ carries
the correlation between $J$ and $P$. 
This is in line with the echo state and FORCE models, where $\mathbf{n}$ corresponds to
the readout vector which is trained and therefore becomes correlated to $J$ and $\mathbf{m}$.
In contrast to these models, however, we constrain the statistics of 
$\mathbf{n}$ to be Gaussian. This allows for an analytical 
treatment and thus for a transparent understanding of how the correlations
affect the network dynamics. 

The details of the construction of $\mathbf{n}$ are described later on. 
At this point we merely state that the 
entries of $\mathbf{n}$ scale with the network size as $1 / N$. 
The structure $P$ is hence considered as a perturbation 
to the random connectivity $J$ whose entries scale as $1 / \sqrt{N}$.
All our results are valid in the limit of infinitely large networks, 
$N \to \infty$. Throughout the work, we compare the theoretical 
predictions with samples from finite networks. 

The network dynamics are given by standard rate equations. Neurons are characterized by their internal 
states $x_i$ and interact with each other via firing rates $\phi(x_i)$. 
The nonlinear transformation from state to firing rate is taken to be 
the hyperbolic tangent, $\phi = \mathrm{tanh}$. 
The entire network dynamics are written as
\begin{equation}
    \label{eq:dot_x}
    \dot{\mathbf{x}}(t) = -\mathbf{x}(t) +  
    \left(J + P
    \right)
    \phi(\mathbf{x}(t))\,,
\end{equation}
with the state vector $\mathbf{x} \in \mathbb{R}^N$ and the 
nonlinearity applied element-wise. 
The derivation of our results, Appendix \ref{sub:mean_field_theory_with_correlations}, 
further includes a constant external input $\mathbf{I}$. 
The results in the main text, however, only consider the autonomous 
network.


\section{Linear dynamics around the origin}%
\label{sec:spectral_properties_of_j}
\begin{figure*}[tb]
    \includegraphics[width=1.\linewidth]{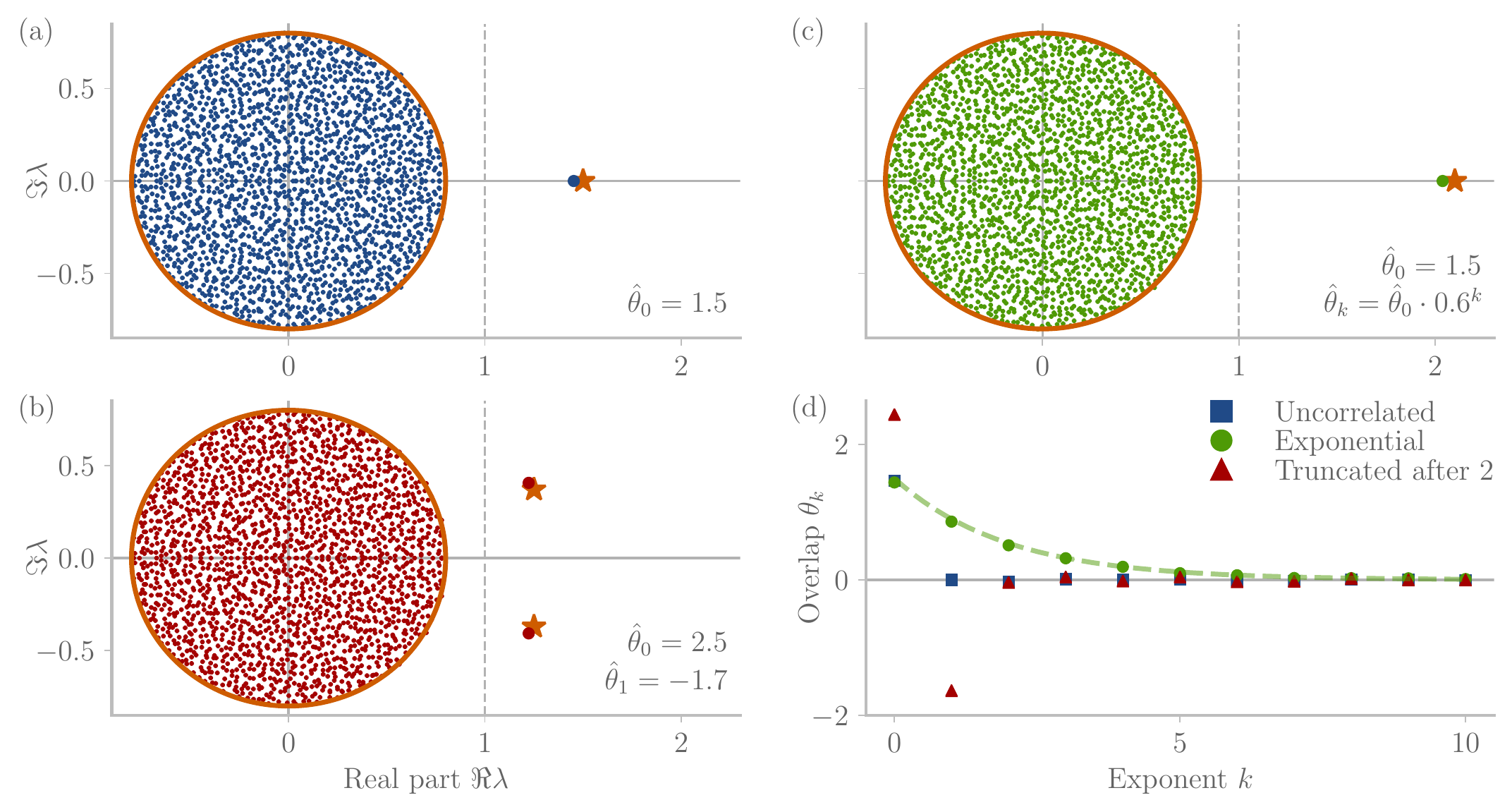}
    \caption{
        Spectral outliers via low-rank perturbations. 
        Spectrum of $J + \mathbf{mn}^T$ with
        (a) no correlations, 
        (b) exponential overlaps, 
        and 
        (c) truncated overlaps, \cref{eq:n_k_2}.
        See \cref{eq:construct_n} 
        for details on the construction of $\mathbf{n}$.
        The values of non-zero $\hat{\theta}_k$ are displayed in
        each plot. 
        Orange circles and stars indicate the theoretical prediction, 
        dots refer to the spectra of the finite-size 
        connectivity matrices, computed numerically. 
        (d) Overlaps $\theta_k = \mathbf{n}^T\! J^k \mathbf{m}$ 
        for the cases above.
        The dashed line are the target overlaps $\hat{\theta}_k$
        for the exponential correlation. 
        Parameters: $N = 2000$, $g = 0.8$.
    }
    \label{fig:outliers_g_08}
\end{figure*}
The origin $\mathbf{x} = \mathbf{0}$ is a fixed point, since $\phi(0) = 0$. 
It is stable if the real parts of all the eigenvalues of the Jacobian are smaller than one. 
Since $\phi'(0) = 1$, the Jacobian is simply the connectivity matrix $J + \mathbf{mn}^T$
itself. Here we examine the spectral properties of this matrix.

\subsection{Eigenvalues}%
\label{sub:eigenvalues}
The spectrum of the Gaussian random matrix $J$ converges to a uniform distribution on
a disk with radius $g$ and centered at the origin for $N \to \infty$ \cite{ginibre1965statistical}.
Previous studies have explored the effect of independent low-rank perturbations 
like in our model \cite{rajan2006eigenvalue, tao2013outliers}.
They found that the limiting distribution of the remaining eigenvalues, 
referred to as the bulk, does not change. Additionally, the spectrum contains 
outliers corresponding to the eigenvalues of the low-rank perturbation itself. 
In this sense, the spectra of the random matrix $J$ and the low-rank perturbation decouple
(although the precise location of each eigenvalue is affected by the perturbation).
To our knowledge, the effect of correlated low-rank perturbations, which we explore below,  
has not been considered before.

To determine the spectrum, we apply the matrix determinant lemma \cite{harville1998matrix}:
\begin{equation}
\det \left(A + \mathbf{mn}^T \right) = 
\left(1 + \mathbf{n}^T\! A^{-1} \mathbf{m}\right) \det(A) \,,
\end{equation}
where  $A \in \mathbb{C}^{N \times N}$ 
is an invertible matrix. For a complex number $z$ that is not an eigenvalue of $J$, 
the matrix $J - \mathds{1} z$ is invertible, resulting in
\begin{equation}
\label{eq:matrix_det_lemma}
\begin{split}
&\det \left((J + \mathbf{mn}^T) - \mathds{1} z \right) 
\\&\qquad= 
\left(1 
+ \mathbf{n}^T\! (J - \mathds{1} z)^{-1} \mathbf{m}\right) \det(J - \mathds{1} z) 
\,.
\end{split}
\end{equation}
The roots of this equation are the eigenvalues of $J + \mathbf{mn}^T$. 
Since the determinant on the right-hand-side is nonzero, we get the scalar equation
\begin{equation}
    \label{eq:eigval_eq}
z = 
\mathbf{n}^T\! \left(\mathds{1} - \frac{J}{z}\right)^{-1} \mathbf{m} \,.
\end{equation}
As long as the entire spectrum is affected by the rank~1 perturbation, 
this equation determines all eigenvalues of $J + \mathbf{mn}^T$. 
We are interested in outliers of the spectrum: eigenvalues of 
$J + \mathbf{mn}^T$ larger than the spectral radius of $J$ (which in the limit of 
$N \to \infty$ is given by $g$). 
For such an outlier, denoted by $\lambda$, the inverse in \cref{eq:eigval_eq} can be
written as a series, and we have
\begin{equation}
    \label{eq:eigval_eq_series}
    \lambda = 
    \sum_{k=0}^\infty \frac{\theta_k}{\lambda^k}
    \,,
\end{equation}
with the overlaps
\begin{equation}
    \label{eq:theta_k}
    \theta_k = \mathbf{n}^T\!J^k\mathbf{m} \,.
\end{equation}
Although this equation is a polynomial of infinite degree, there can be at most proper $N$ solutions (those outside of the bulk; see Appendix \ref{sub:finite_solutions}). 

The series representation \cref{eq:eigval_eq_series} is the main result of this section.
It indicates that the overlaps $\theta_k$ between $\mathbf{m}$ and $\mathbf{n}$ 
after passing through $J$ for $k$ times determine the eigenvalues of 
the perturbed matrix. It is hence useful to characterize the correlations between 
$J$ and the rank-one perturbation in terms of these overlaps. 

The description up to this point is general and does not depend on details of the matrix $J$. 
For our model, where $J$ is a random matrix, the scalar products
$\theta_k$ over $N$ entries are self-averaging: for $N \to \infty$, $\theta_k$ converges to its ensemble average $\mathbb{E}[\theta_k]$, with variance decaying as $1 / N$. 
We rely on this property and compute quantities for single realization of large networks instead of ensemble averages. 

A random matrix $J$ has the effect of decorrelating independent vectors:
if the vectors  $\mathbf{m}$ and $\mathbf{n}$ are
uncorrelated to $J$, a single pass through the network already annihilates any 
overlap between $\mathbf{n}$ and $J\mathbf{m}$.
In Appendix \ref{sub:construction_of_the_vector_n}, we formally show that the self-averaging indeed
yields $\mathbf{n}^T\!J^k\mathbf{m}=0$ for $k \ge 1$.
We can apply this to 
\cref{eq:eigval_eq_series}: If $\theta_k = 0$ for any $k \ge 1$, then 
\begin{equation}
    \lambda = \theta_0 = \mathbf{n}^T\!\mathbf{m}\,.
\end{equation}
Thus an independent rank-one perturbation yields a single outlier positioned at the 
eigenvalue of the rank-one matrix itself [\cref{fig:outliers_g_08}(a)], in accordance with
known results \cite{rajan2006eigenvalue, tao2013outliers}.

If $\mathbf{mn}^T$ is correlated to $J$, the $\theta_k$ will not vanish for nonzero $k$. 
We analyze two special cases:
\begin{enumerate}[label=(\roman*)]
    \item 
        If $\theta_k = 0$ for any $k \ge 2$, then
        there are two outliers 
        \begin{equation}
            \label{eq:lambda_pm}
            \lambda_\pm = 
            \frac{\theta_0}{2} \pm
            \sqrt{
                \left(\frac{\theta_0}{2}\right)^2 
                +
                \theta_1
            } \,.
        \end{equation}
        This can give rise to complex conjugate outliers, as
        displayed in \cref{fig:outliers_g_08}(b).
        More generally, $K$ nonzero overlaps lead to $K$ outliers
        via a polynomial equation
        [\cref{eq:lambda_poly_trunc}].
    \item 
        A second case is one of a converging series in \cref{eq:eigval_eq_series}. 
        The simplest assumption is an exponential scaling, $\theta_k = \theta_0 b^k$
        with base $b$. Inserting into the eigenvalue 
        equation \eqref{eq:eigval_eq_series}
        yields a single solution
        \begin{equation}
            \lambda = \theta_0 + b \,.
        \end{equation}
        Remarkably, we see that correlation between the random matrix $J$ 
        and the rank-one perturbation does not necessarily lead to more than one 
        outlier.
        This is shown in \cref{fig:outliers_g_08}(c).
        The observation generalizes to correlations expressible as a sum of $K$
        exponentially decaying terms, leading to $K$ outliers
        [\cref{eq:lambda_poly_exp}].
\end{enumerate}
We can apply this understanding to construct a network with a set of outliers 
and either one of the underlying correlation structures. 
One way is to define the vector $\mathbf{n}$ explicitly in terms of $\mathbf{m}$ 
and $J$. 
For example, if we set
\begin{equation}
    \label{eq:n_k_2}
    \mathbf{n} = 
    \frac{1}{N} \left(
    \hat{\theta}_0 \, \mathbf{m} +
    \frac{\hat{\theta}_1}{g^2} J \mathbf{m} 
\right) \,,
\end{equation}
then the overlaps will self-average to $\mathbb{E}[\theta_k] = \hat{\theta}_k$ 
for $k \in \{0, 1\}$ and $\mathbb{E}[\theta_k] = 0$ for any $k \ge 2$, with variance 
scaling as $1 / N$. 
This is shown formally and generalized to higher $\theta_k$
in Appendix \ref{sub:construction_of_the_vector_n}. The details of the construction 
for a set of target outliers is further detailed in Appendix \ref{sub:construction_of_outliers}.
The discrepancy between numerical and target outliers in Figure \ref{fig:outliers_g_08} is due to finite size effects, which decay with $1 / \sqrt{N}$ (verified numerically, and in accordance with Ref. \cite{mastrogiuseppe2018linking}). 

The simulations further show that the remaining eigenvalues span the same circle
as without the perturbation. 
While all eigenvalues change, visual inspection does not reveal any changes in the statistics.

\subsection{Implementation of multiple outliers}
\label{sub:implementation_of_multiple_outliers}
\begin{figure}[tb]
    \includegraphics[width=1.\linewidth]{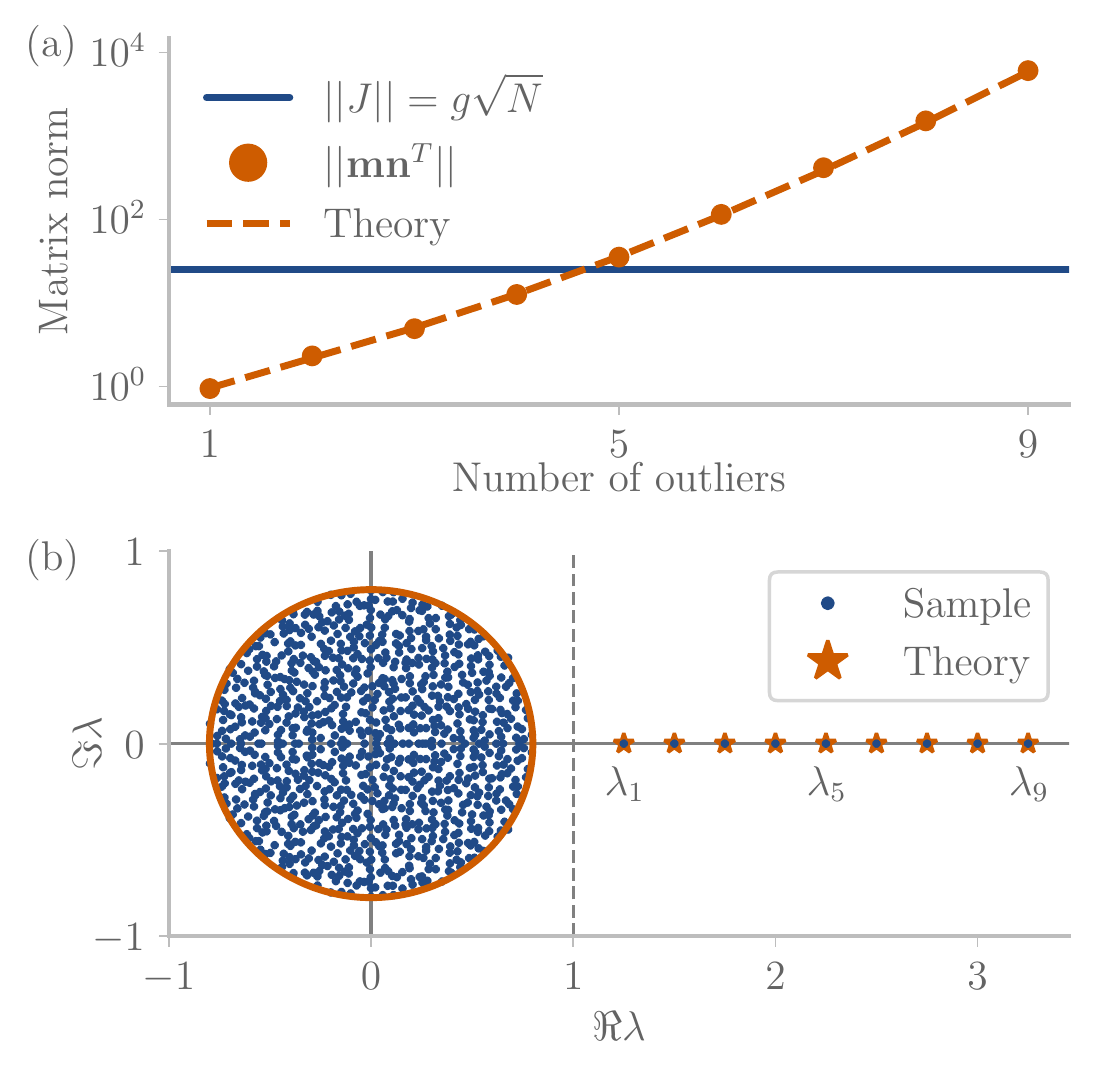}
    \caption{Scaling of the norm of the rank-one perturbation with number of induced outliers.
        The vector $\mathbf{n}$ is the least square solution to implementing 
        a set of outliers $\Lambda = \{\lambda_1, \dots,\lambda_K\}$, 
        with 
        $\lambda_k = 1.25 + 0.25 k$,
        see Appendix \ref{sub:least_square_vector_n}.
        (a) Log-linear plot of the Frobenius norm of $J$ and $\mathbf{mn}^T$ 
        as a function of the number of outliers.
        The dashed line is the  theoretical prediction.
        (b) Spectrum of $J + \mathbf{mn}^T$ for $K=9$ outliers. 
        Parameters: $N = 1000$, $g = 0.8$.
    }
    \label{fig:scaling_with_outliers_N_1000_g_08}
\end{figure}

So far we analyzed the outliers for given correlations between $J$ 
and $\mathbf{mn}^T$ as quantified by the overlaps $\theta_k$. 
We now change the perspective and ask about the properties of the 
rank-one perturbation given a set of outliers. 
We saw that in principle a given set of outliers  may have multiple 
underlying correlation structures -- e.g. through a truncated set 
of non-zero overlaps or a combination of exponentially decaying terms. 
Regardless of the correlation structure, however, we observe that 
the norm of $\mathbf{n}$ grows fast with the number of outliers introduced, 
implying that strong perturbations are needed to generate a large number of 
outliers.

To understand analytically the origin of this phenomenon, we focus on a 
method to determine the least square $\mathbf{n}$
given $J$, $\mathbf{m}$ and the 
set of target outliers $\Lambda$. 
The resulting $\mathbf{n}$
can be formulated using the pseudoinverse, as detailed in
Appendix \ref{sub:least_square_vector_n}. 
The main result of this analysis is the scaling of the Frobenius norm of 
the rank-one matrix $\mathbf{mn}^T$ with the number of outliers. 
The asymptotic behavior is given by
\begin{equation}
    \label{eq:scaling_mn}
    ||\mathbf{mn}^T||
    \sim g \prod_{\lambda \in \Lambda} 
    \frac{|\lambda|}{g} \,,
\end{equation}
that is, exponentially growing with the number of outliers. 
In comparison, the Frobenius norm of $J$ is given by $||J|| = g \sqrt{N}$. 
This means that if one aims to place more than a handful of outliers, 
the perturbation $\mathbf{mn}^T$ becomes 
the dominating term (for a fixed network size $N$). We illustrate this in 
\cref{fig:scaling_with_outliers_N_1000_g_08} by plotting $||\mathbf{mn}^T||$ 
for sets of outliers $\Lambda_K = \{\lambda_1, \dots, \lambda_K\}$ 
with growing number $K$. The outliers $\lambda_k$ were placed on the real line.
Further tests including complex eigenvalues gave similar
results (not shown).
A similar method of deriving $\mathbf{n}$ from the pseudoinverse has been described in
Ref. \cite{logiaco2019model}.

The scaling \eqref{eq:scaling_mn} shows another important point: the bulk radius
$g$ critically determines the norm of the rank-one perturbation. Indeed, the 
contribution of each outlier $\lambda_k$ is relative to the radius. 
Even for a single outlier, where
\begin{equation}
    ||\mathbf{mn}^T|| = \sqrt{\lambda^2 - g^2} \,,
\end{equation}
an increase in $g$ leads to decreasing norm. 
This observation suggests that a large random connectivity
facilitates the control of the spectrum by a rank-one perturbation.

\section{Non-trivial fixed points}%
\label{sec:non_trivial_fixed_points}
We now turn to the non-trivial fixed points of the network.
At these, the internal states $\mathbf{x}$ obey the equation
\begin{equation}
    \label{eq:fixedpoint_rank1}
    \mathbf{x} = J \bm{\phi} + \kappa \mathbf{m}\,.
\end{equation}
Here we defined the scalar feedback strength $\kappa = \mathbf{n}^T\! \bm{\phi}$, 
using the vector notation $\bm{\phi} = \phi(\mathbf{x})$. 

The fixed points of related models have been analyzed in previous 
works. For infinitely large networks, the unperturbed system ($P = 0$)
has a single fixed point at the origin if $g < 1$ \cite{wainrib2013topological}.
For $g > 1$, the system exhibits chaotic dynamics \cite{sompolinsky1988chaos}.
In this regime, the number of (unstable) fixed points scales exponentially with the network size 
$N$ \cite{wainrib2013topological}.
Here we only focus on networks in the non-chaotic regime, where either $g < 1$ 
or the perturbation $P$ suppresses chaos 
\cite{mastrogiuseppe2018linking}.

\subsection{Fixed point manifold}%
\label{sub:fixed_point_manifold}
\begin{figure*}[tb]
    \includegraphics[width=1.\linewidth]{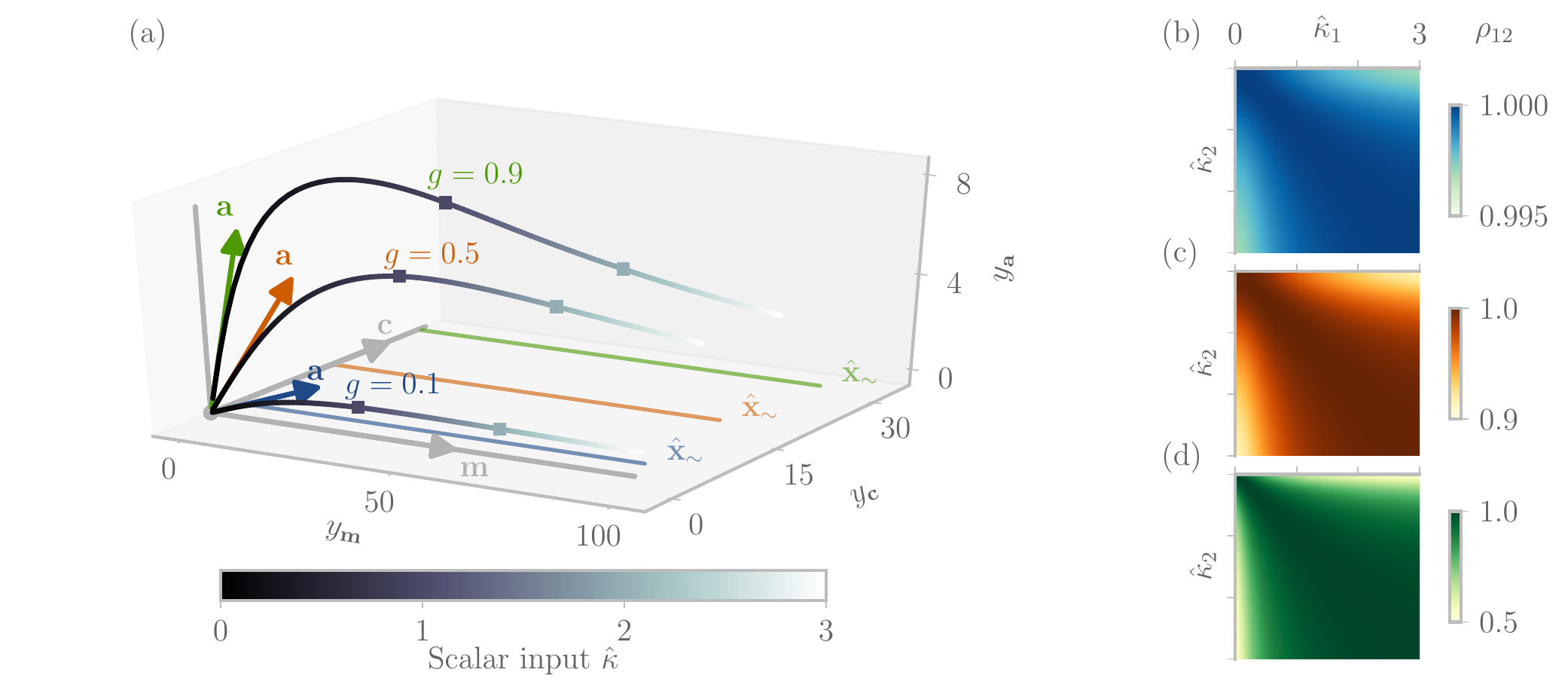}
    \caption{Manifold $\mathcal{M}$, \cref{eq:M} constraining fixed points, 
        \cref{eq:x_fp_ol}.
        (a)
        Projection of $\mathcal{M}$ for three networks with 
        different strength of randomness $g$ (see main text for the three-dimensional basis).
        The negative side for $\hat{\kappa} < 0$ is symmetric and not shown. 
        The squares on the manifolds indicate the inputs $
        \hat{\kappa}= (1, 2)$.
        The straight lines in the plane $y_\mathbf{a} = 0$ are the asymptotic directions 
        $\hat{\mathbf{x}}_\sim$
        for the manifolds.
        (b-d)
        Correlation $\rho_{12}$ between two points $\hat{\mathbf{x}}(\hat{\kappa}_i)$
        on $\mathcal{M}$ for two different inputs $\hat{\kappa}_1, \hat{\kappa}_2 \in [0, 3]$. 
        (b, c, d) correspond to the random connectivity 
        strengths $g = 0.1, 0.5, 0.9$, respectively. 
        Note the different scales on the color bars.
    }
    \label{fig:open_loop_fp_manifold}
\end{figure*}

Following \citet{rivkind2017local}, the perturbed system with fixed points 
\eqref{eq:fixedpoint_rank1} can be understood using a surrogate system in which 
the feedback $\kappa$ is replaced with a fixed scalar $\hat{\kappa}$.
For $g < 1$, every such value $\hat{\kappa}$ corresponds to a unique
fixed point 
\begin{equation}
    \label{eq:x_fp_ol}
    \hat{\mathbf{x}} = J \phi(\hat{\mathbf{x}}) + \hat{\kappa} \mathbf{m}\,.
\end{equation}
This equation defines the one-dimensional nonlinear manifold 
\begin{equation}
    \label{eq:M}
    \mathcal{M} = \{\hat{\mathbf{x}} \,|\, \hat{\kappa} \in \mathbb{R}\} \,.
\end{equation}

The manifold $\mathcal{M}$ can be understood by looking at the
asymptotics.
For large input $\hat{\kappa}$, the nonlinearity saturates and the manifold becomes 
approximately linear:
\begin{equation}
    \label{eq:x_fp_ol_asymp}
    \hat{\mathbf{x}}_\sim = \mathbf{c} + \hat{\kappa} \mathbf{m} \,,
\end{equation}
with $\mathbf{c} = J\, \mathrm{sign}(\mathbf{m})$. 
Around the origin, we linearize and obtain
\begin{equation}
    \label{eq:x_fp_ol_orig}
    \hat{\mathbf{x}} = \hat{\kappa} \,\mathbf{a} + \mathcal{O}(\hat{\kappa}^2)\,,
\end{equation}
with $\mathbf{a} = \left( \mathds{1} - J\right)^{-1} \mathbf{m}$.

Applying orthonormalization to the triplet  
$(\mathbf{m, c, a})$, we obtain a three-dimensional basis.
We observe that, for $N \to \infty$, the vectors $\mathbf{m}$ and $\mathbf{c}$ are
orthogonal and that the vector $\mathbf{a}$ becomes orthogonal to the other two
in the limit $g \to 1$. Accordingly, we name the coefficients of the basis
$(y_\mathbf{m}, y_\mathbf{c}, y_\mathbf{a})$.
The projection of the manifold $\mathcal{M}$ on this basis is shown 
in \cref{fig:open_loop_fp_manifold}(c) for three different values of $g$. 
Numerical evaluation of the reconstruction error shows that these three 
dimensions reconstruct the manifold very well
albeit with decreasing accuracy for increasing $g$ (not shown).

Fixed points of the full system are obtained by determining $\kappa$ 
self-consistently. They necessarily 
lie on the manifold $\mathcal{M}$. One consequence is a strong correlation
between pairs of fixed points, especially if both lie close to the origin or 
in the saturating regime. 
In \cref{fig:open_loop_fp_manifold}(b-d), we numerically evaluate
this correlation for three different randomness strengths $g$.
On can observe that for $g \le 0.5$, correlation does not drop below 90\%.
Even for $g = 0.9$, the correlation is low only if one fixed point is very close to the origin 
and the other one is far out.

So far we only considered the case $g < 1$. 
For $g > 1$, there is a minimal $\hat{\kappa}_\mathrm{min}$ for which 
the dynamics are stabilized and a unique stable fixed point emerges
\cite{mastrogiuseppe2018linking}.
Here, the manifold $\mathcal{M}$ is unconnected and now reads 
$\mathcal{M} = \{\hat{\mathbf{x}} \,|\, 
\hat{\kappa} \in \mathbb{R} \setminus (-\hat{\kappa}_\mathrm{min}, \hat{\kappa}_\mathrm{min}) \}$.%

Finally we note that the constraints of a one-dimensional manifold are general 
and do not depend on the details of the vector $\mathbf{n}$, especially not on its Gaussian statistics. 
This is particularly important for learning algorithms
like the echo state framework or FORCE, which by construction only allow for the adaptation of 
the vector $\mathbf{n}$
\cite{jaeger2004harnessing, sussillo2009generating}.
Accordingly, fixed points in these cases are also
strongly correlated, which may lead to catastrophic forgetting when trying
to learn multiple fixed points sequentially \cite{beer2019one}.

\subsection{Mean field theory}%
\label{sub:mean_field_theory}
\begin{figure*}[tb]
    \includegraphics[width=1.\linewidth]{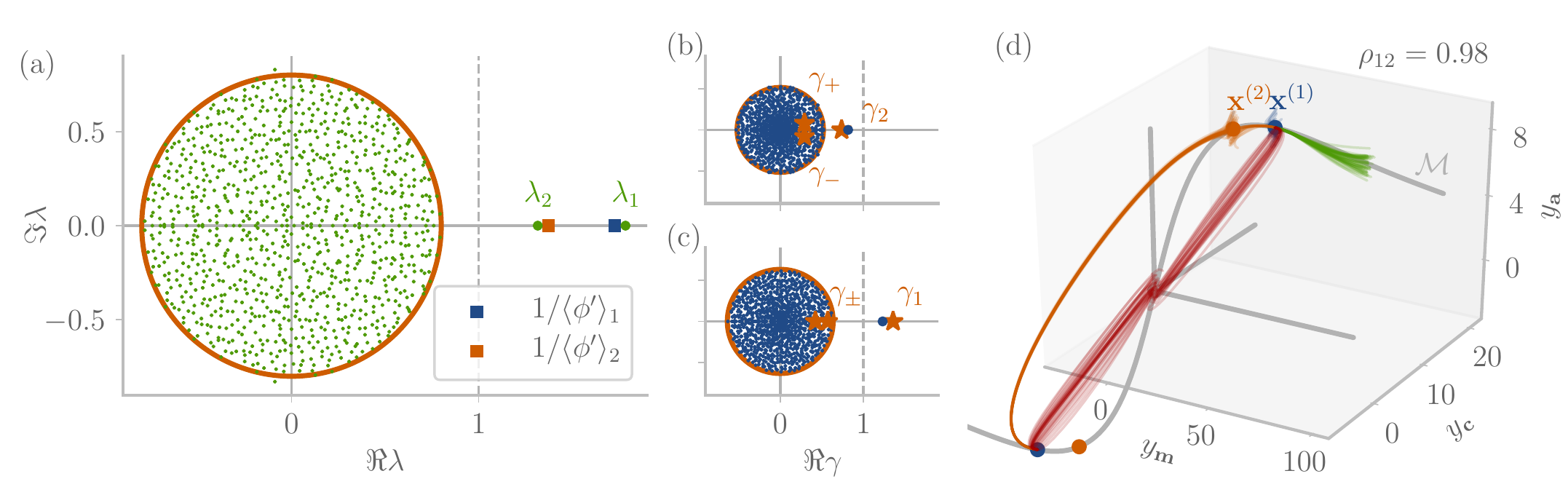}
    \caption{Two fixed points induced by rank-one perturbation correlated to
        the random connectivity $J$. 
        (a) Spectrum of  $J + \mathbf{mn}^T$. The squares indicate the corresponding
        averaged slopes at the fixed points, as predicted by \cref{eq:eigval_dphi}. 
        (b-c) Eigenvalues of the Jacobian at the two fixed points $\mathbf{x}^{(1)}$ (b)
        and $\mathbf{x}^{(2)}$ (c).
        Stars indicate the theoretical predictions for exceptional 
        stability eigenvalues (only meaningful outside the bulk). 
        (d)
        Fixed points and manifold $\mathcal{M}$. 
        The colored lines indicate trajectories starting from the two fixed points
        (blue and orange), a point on the manifold $\mathcal{M}$ (green) and the origin.
        All trajectories converge on $\mathbf{x}^{(1)}$ or its negative counterpart.
        At each point, 50 different initial conditions are obtained by adding 
        Gaussian noise (SD = 0.5).
        The fixed point correlation is indicated by $\rho_{12}$.
        Parameters: $g = 0.8$, $N = 1000$.
        The rank-one perturbation is obtained by 
        the least-squared $\mathbf{n}$, see Appendix \ref{sub:least_square_vector_n}.
    }
    \label{fig:dynamics_2fps_rank1_corr_g_08}
\end{figure*}

For non-trivial fixed points of the full network, \cref{eq:fixedpoint_rank1},
the scalar feedback $\kappa$ 
needs to be consistent with the firing rates $\phi(\mathbf{x})$. 
Similar to prior works, we compute $\kappa$ using a mean field theory
\cite{mastrogiuseppe2018linking}. The central idea of the mean field theory
is to replace the input to each variable $x_i$ by a stochastic variable with statistics
matching the original system. The statistics of the resulting stochastic processes
$x_i$ are then computed self-consistently. 

Because our model includes correlations between the random part $J$ and the 
low-rank structure $P$, the correlations in the activity do not vanish as dynamics unfold.
This phenomenon prevents the application of previous  
theories \cite{mastrogiuseppe2018linking}. We hence develop a new theory. 
The details are elaborated in
Appendix \ref{sub:mean_field_theory_with_correlations}.
Here we give an outline of the analysis. 

The starting point is the scalar 
feedback $\kappa$. The Gaussian statistics of 
$\mathbf{n}$ and the fixed point $\mathbf{x}$ allow to factor out the effect 
of the nonlinearity via partial integration. We have
\begin{equation}
    \label{eq:kappa_partial}
    \kappa 
    = \mathbf{n}^T\!\bm{\phi} 
    = \langle \phi' \rangle \,\mathbf{n}^T\! \mathbf{x} \,,
\end{equation}
with the average slope $\langle\phi'\rangle$ evaluated at the fixed point:
\begin{equation}
    \langle \phi' \rangle
    = \int \mathcal{D}z \,\phi'(\sqrt{\Delta^0} z)  \,,
\end{equation}
where $\mathcal{D}z$ is the standard Gaussian measure. $\Delta^0$ is the variance of $\mathbf{x}$, which from the fixed point equation 
\eqref{eq:fixedpoint_rank1} is given by
\begin{equation}
    \Delta^0 
    = g^2 \langle \phi^2 \rangle 
    + \kappa^2 
    \,.
\end{equation}
The quantities $\langle\phi'\rangle$, $\Delta^0$, and $\kappa$ are determined self-consistently. 
To that end, we further evaluate $\kappa$ in \cref{eq:kappa_partial}.
Inserting the fixed point equation 
\eqref{eq:fixedpoint_rank1}
yields
\begin{equation}
    \label{eq:insert_x}
    \mathbf{n}^T\! \mathbf{x} 
    =
    \mathbf{n}^T\! J \bm{\phi}
    + \kappa \mathbf{n}^T\! \mathbf{m} \,.
\end{equation}
The first term on the right-hand side vanished in previous 
studies with no correlation between $P$ and $J$ \cite{mastrogiuseppe2018linking}. In our case, there are correlations, and we proceed to analyze this term. We first  interpret $J^T\mathbf{n}$ as 
a Gaussian vector and use partial integration to replace $\bm{\phi}$ with $\mathbf{x}$:

\begin{equation}
    \mathbf{n}^T\! \mathbf{x} 
    = \langle \phi' \rangle \mathbf{n}^T\! J \mathbf{x} 
    + \kappa \mathbf{n}^T\! \mathbf{m} \,.
\end{equation}

We now insert the fixed point equation \eqref{eq:fixedpoint_rank1} into the new first term of the right hand side.
We can apply this scheme recursively and arrive at an equation 
linear in $\kappa$ on both sides:
\begin{equation}
    \kappa
    = \kappa \langle\phi'\rangle \,
    \sum_{k=0}^\infty \langle \phi' \rangle^k \,\theta_k
    \,,
\end{equation}
with overlaps as defined above, \cref{eq:theta_k}.
We are looking at a non-trivial fixed point, so we can divide by the 
nonzero $\kappa$ to obtain
\begin{equation}
    \label{eq:dphi}
    \frac{1}{\langle \phi' \rangle} = 
    \sum_{k=0}^\infty \langle \phi' \rangle^k \,\theta_k
    \,.
\end{equation}
A comparison with \cref{eq:eigval_eq_series} shows that the two
equations are identical if
\begin{equation}
    \label{eq:eigval_dphi}
    \lambda = 1 / \langle \phi' \rangle \,. 
\end{equation}  
This is a remarkable relationship between the outliers and 
autonomously generated fixed points:
each non-trivial fixed point $\mathbf{x}^{(i)}$ must be associated with 
a real eigenvalue $\lambda_i$ such that the average over the derivative
of firing rates at this fixed point, $\langle \phi'\rangle_i$,
fulfills the above condition \eqref{eq:eigval_dphi}.
In the special case of $\phi = \mathrm{tanh}$, 
the $\langle \phi'\rangle_i$ are confined to the interval $(0, 1]$, 
so the corresponding eigenvalues must be real and larger than one. 
One may hence look at the spectrum of the connectivity matrix alone and determine
how many non-trivial fixed points there are. 
An instance of this phenomenon is illustrated in 
\cref{fig:dynamics_2fps_rank1_corr_g_08}. The spectrum at the origin 
contains two outliers $\lambda_i$, $i=1, 2$, each real and larger than one. 
The dynamics have two corresponding fixed points $\mathbf{x}^{(i)}$
located on the manifold $\mathcal{M}$. 
In accordance with \cref{eq:eigval_dphi}, the average slopes at these fixed points, 
$1 / \langle\phi'\rangle_i$, agree with the outliers up to deviations due to the finite network size. 

\subsection{Stability of fixed points}%
\label{sub:stability_of_fixed_points}
\begin{figure*}[tb]
    \includegraphics[width=1.\linewidth]{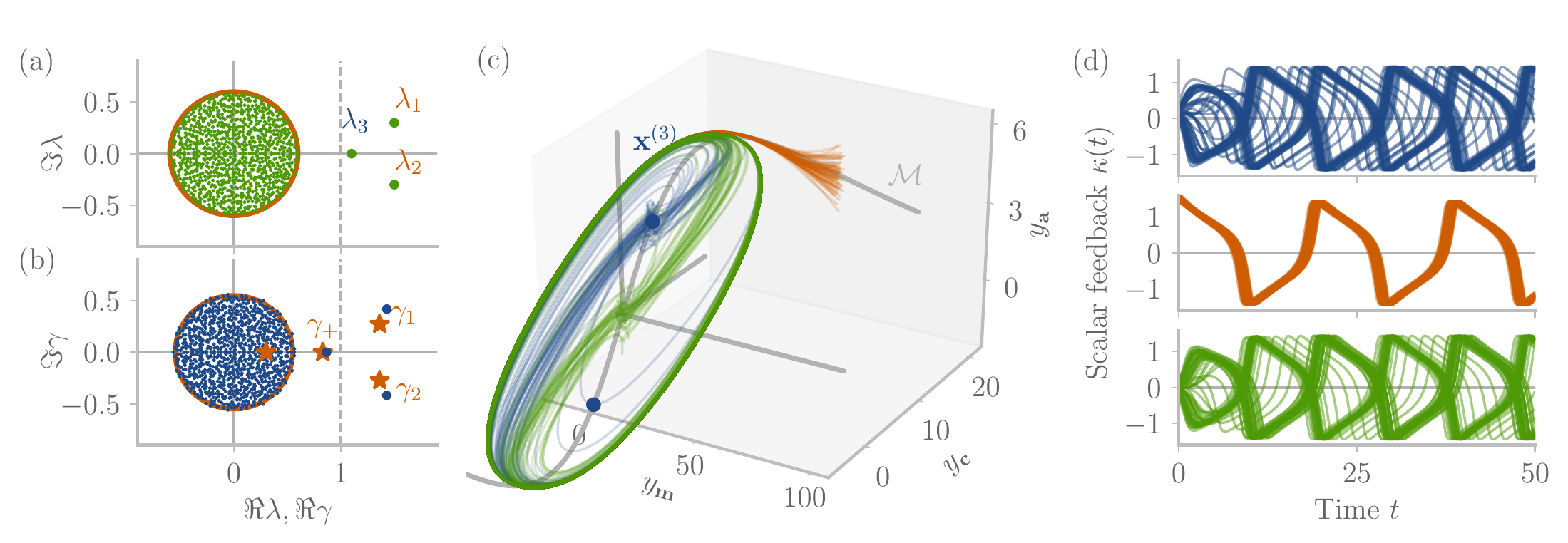}
    \caption{Limit cycle induced by oscillatory instability. 
        (a) Spectrum of the connectivity matrix. The outlier $\lambda_3$ 
        is real-valued and larger than one, so there is a corresponding fixed point
        $\mathbf{x}_3$. 
        (b) The spectrum at the fixed point. The 
        predicted stability eigenvalue $\gamma_-$ lies inside the bulk and is not labeled.
        (c) Fixed point and dynamics
        as in \cref{fig:dynamics_2fps_rank1_corr_g_08}.
        Trajectories start at the fixed point (blue), a point on the 
        manifold $\mathcal{M}$ (orange) and the origin (green).
        (d) Scalar feedback $\kappa(t)$ for the different initial
        conditions. 
        Parameters: $g = 0.6$, $N = 1000$.
        The rank-one perturbation is obtained by 
        the least-squared $\mathbf{n}$, see Appendix \ref{sub:least_square_vector_n}.
    }
    \label{fig:oscillations_rank1_corr_g_06}
\end{figure*}
The stability of each fixed point is determined by the spectrum of its Jacobian.
The associated stability matrix (the Jacobian shifted by $-\mathds{1}$)
is 
\begin{equation}
    S = \left( J + \mathbf{m}\mathbf{n} \right) R' \,,
\end{equation}
with the diagonal matrix of slopes $R_{ij}' = \delta_{ij} \phi_i'$.
Previous work (\cite{mastrogiuseppe2018linking}) has found that the spectrum of $S$, too, 
consists of a bulk and a small number of exceptional eigenvalues: 
in the case of an uncorrelated rank-one perturbation, 
there are two nonzero eigenvalues obtained via mean field theory, only one of which 
has been found outside the random bulk.
The radius of the bulk shrinks to $g \sqrt{\langle \phi'^2\rangle}$
due to the saturation of the nonlinearity \cite{mastrogiuseppe2018linking}.
We find numerically that the bulk behaves alike in our model, too. For the rest of this section,
however, we focus on exceptional eigenvalues of the stability matrix $S$, denoted by $\gamma$. 

Similar to the trivial fixed point, one can apply the matrix determinant lemma to 
derive an equation for the stability eigenvalues:
\begin{equation}
    \label{eq:eigval_eq_gamma}
\gamma = 
\mathbf{n}^T\! R' \left(\mathds{1} - \frac{J R'}{\gamma}\right)^{-1} \mathbf{m} 
\,.
\end{equation}
We can apply the mean field theory introduced above to evaluate the right hand side. 
The details of this calculation are deferred to Appendix
\ref{sub:stability_eigenvalues}. It turns out that 
the resulting $\gamma$ are surprisingly compact. We now describe these 
stability eigenvalues. 

Consider the fixed point $\mathbf{x}^{(i)}$. 
According to \cref{eq:eigval_dphi},
this fixed point corresponds to the eigenvalue $\lambda_i$. 
\cref{eq:eigval_eq_gamma} always has two solutions
$\gamma_\pm$ determined by a quadratic equation. 
These are only dependent on the outlier $\lambda_i$ and the 
statistics of the fixed point $\mathbf{x}^{(i)}$, but entirely
independent of the remaining spectrum or other fixed points. 
Their precise values are detailed in  
\cref{eq:gamma_pm}. It turns out that $\gamma_+$ and $\gamma_-$ always 
have a real part smaller than one. They hence do not destabilize the fixed point. 
Additionally, at least one of the two is always hidden within the bulk of the 
eigenvalues, as observed before for the case of no correlation
\cite{mastrogiuseppe2018linking}.
In \cref{fig:dynamics_2fps_rank1_corr_g_08}(b-c), the spectra of the Jacobian 
at two fixed points are compared with the theoretical predictions. In both 
cases, $\gamma_\pm$ correspond to the two stars within the bulk.
In \cref{fig:oscillations_rank1_corr_g_06}(b), the bulk is smaller ($g = 0.6$) and 
$\gamma_+$ is visible. 

If $\lambda_i = \lambda_1$ is the only outlier, then $\gamma_\pm$ are the only two
solutions of \cref{eq:eigval_eq_gamma}, and the fixed point
$\mathbf{x}^{(1)}$ as well as its mirror $-\mathbf{x}^{(1)}$ will be stable. 
However, if there are $K \ge 2$ outliers $\{\lambda_1, \dots, \lambda_K\}$, 
we find an additional set of $K - 1$ stability eigenvalues
\begin{equation}
    \label{eq:gamma_j}
    \gamma_j = \frac{\lambda_j}{\lambda_i} \quad \text{for all} \quad 
    j \in \{1, \dots, K\}, \, j \ne i \,.
\end{equation}
This expression indicates a remarkable relationship between different fixed points:
the existence of a fixed point $\mathbf{x}^{(j)}$ 
with outlier $\lambda_j > \lambda_i$ will always 
destabilize the fixed point $\mathbf{x}^{(i)}$ corresponding to $\lambda_i$. 
Conversely, if there are no outliers with real part larger than
that of $\lambda_i$, then $\mathbf{x}^{(i)}$ will be stable. 
Since 
$\lambda = 1 / \langle\phi'\rangle$ implies that larger $\lambda$ corresponds
to larger fixed point variance $\Delta^0$, one can say that
only the largest fixed point can be stable, 
Such an interaction between two fixed points is illustrated in 
\cref{fig:dynamics_2fps_rank1_corr_g_08}(b-c). The stars outside of the bulk 
correspond
to the predicted eigenvalue $\gamma_2$ or $\gamma_1$. Comparison between the
theoretical prediction \eqref{eq:gamma_j} and numerical calculation for a 
sampled network shows good agreement for both fixed points. 
Furthermore, a simulation of the dynamics in \cref{fig:dynamics_2fps_rank1_corr_g_08}(d)
shows that indeed all trajectories converge to the larger fixed point $\mathbf{x}^{(1)}$
or its negative counterpart. 

Finally, note that a complex outlier $\lambda_j$
also destabilizes a fixed point $\mathbf{x}^{(i)}$ if the real part of $\lambda_j$
is larger than that of $\lambda_i$. 
Complex outliers do not have corresponding fixed points, since \cref{eq:eigval_dphi} is real. 
An example of such a case is shown in \cref{fig:oscillations_rank1_corr_g_06}. 
There is only one real eigenvalue larger than one, 
and hence only a single non-trivial fixed point $\mathbf{x}^{(3)}$. 
Nonetheless, the two complex outliers $\lambda_1 = \lambda_2^*$ destabilize the fixed point
by virtue of \cref{eq:gamma_j}, since the real parts are larger than the outlier
corresponding to the fixed point, $\Re\lambda_1 = \Re\lambda_2 > \lambda_3$.
Numerical simulations indicate that in such a case, the dynamics converge on a limit cycle. 

\section{rank-two perturbations}%
\label{sec:rank_2}
\begin{figure*}[tb]
    \includegraphics[width=1.\linewidth]{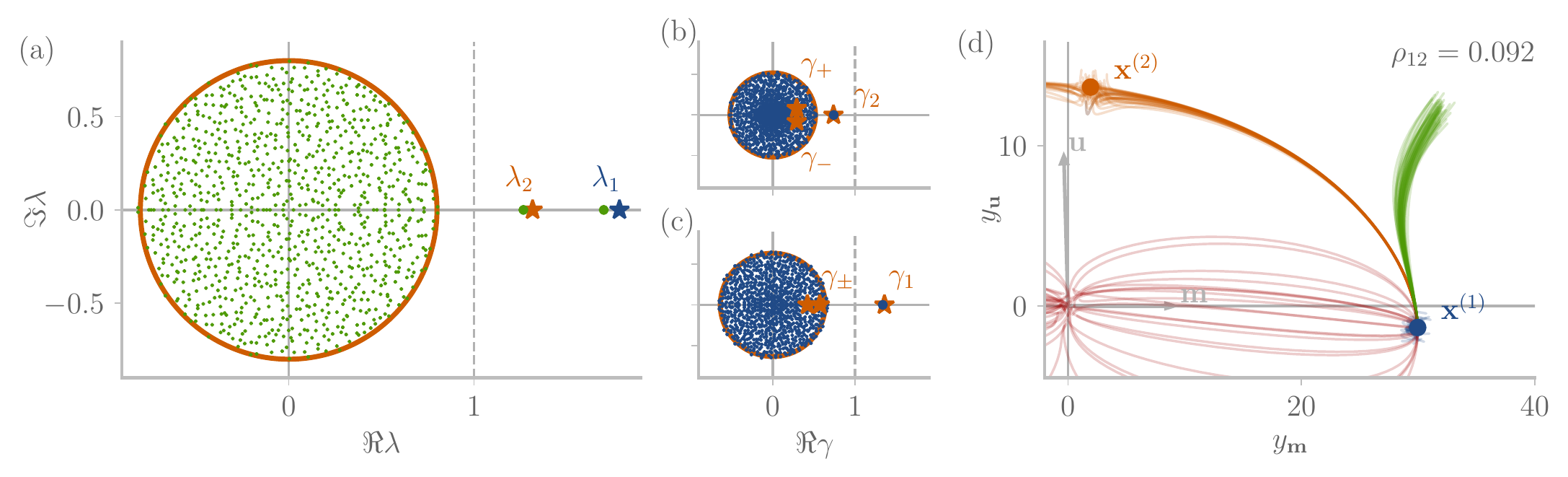}
    \caption{Fixed points and dynamics for a rank-two perturbation with
        structures $\mathbf{mn}^T$ and $\mathbf{uv}^T$ drawn independently of each other 
        as well as of $J$.
        (a-c) Spectra of the Jacobian at the origin (a)
        and at the two fixed points $\mathbf{x}^{(1)}$ (b)
        and $\mathbf{x}^{(2)}$ (c).
		Stars denote the predictions for infinite size networks. 
        (d) Projection of fixed points on vectors $\mathbf{m}$ and $\mathbf{u}$, 
        and trajectories starting around $\mathbf{x}^{(1)}$ (blue),
        $\mathbf{x}^{(2)}$ (orange), $\mathbf{x}^{(1)} + \mathbf{x}^{(2)}$ (green) and $\mathbf{0}$ (red).
        The correlation between the two fixed points is indicated by $\rho_{12}$.
        Other parameters as in \cref{fig:dynamics_2fps_rank1_corr_g_08}.
    }
    \label{fig:dynamics_2fps_uncorr_g_08}
\end{figure*}

The previous section demonstrated two properties of networks with multiple non-trivial fixed points: they are highly correlated due to the confinement on the manifold 
$\mathcal{M}$ [\cref{fig:open_loop_fp_manifold}(b-d) and
\cref{fig:dynamics_2fps_rank1_corr_g_08}(d)], and their stability properties interact [\cref{eq:gamma_j}]. 
We asked whether the latter is a result of the former.
To approach this question, we extend the model to a rank-two perturbation
which allows for uncorrelated fixed points.

The rank-two connectivity structure is defined by 
\begin{equation}
    \label{eq:P_rank2}
    P = \mathbf{m} \mathbf{n}^T + \mathbf{u} \mathbf{v}^T \,.
\end{equation}
We assume $J$, $\mathbf{m}$ and $\mathbf{u}$ to be drawn independently. 
Similar to the rank-one case, the entries of both $\mathbf{m}$ and $\mathbf{u}$ are drawn from 
standard normal distributions
while $\mathbf{n}$ and $\mathbf{v}$ are Gaussian and dependent on $J, \mathbf{m}$ and $\mathbf{u}$.

The outliers $\lambda$ of the perturbed matrix 
$J + \mathbf{mn}^T\! + \mathbf{uv}^T$ are calculated similarly to the rank-one case. 
Applying the matrix determinant lemma twice, we arrive at an equation 
of quadratic form:
\begin{equation}
    \label{eq:eigvals_r2}
    0 = \lambda^2 - \lambda \mathrm{Tr} Q_\lambda + \det(Q_\lambda) \,.
\end{equation}
In other words, $\lambda$ is an eigenvalue of the matrix 
\begin{equation}
    Q_\lambda = 
    \begin{bmatrix}
        \mathbf{n}^T\! M_\lambda \mathbf{m} &
        \mathbf{n}^T\! M_\lambda \mathbf{u} \\
        \mathbf{v}^T\! M_\lambda \mathbf{m} &
        \mathbf{v}^T\! M_\lambda \mathbf{u}
    \end{bmatrix} \,,
\end{equation}
which depends on $\lambda$ through 
\begin{equation}
    \label{eq:M_lam}
    M_\lambda = 
    \left(\mathds{1} - \frac{J}{\lambda}\right)^{-1} \,.
\end{equation}
In general there are more than two solutions, but 
if the rank-two perturbation is uncorrelated to $J$, the matrix $M_\lambda$ 
disappears in $Q_\lambda$. The solution is then
in agreement with previous results
\cite{mastrogiuseppe2018linking}.

Non-trivial fixed points of the network dynamics \eqref{eq:dot_x} 
with a rank-two perturbation \eqref{eq:P_rank2}
obey the equation
\begin{equation}
    \label{eq:fixedpoint_rank2}
    \mathbf{x} = J \bm{\phi} 
    + \kappa_1 \mathbf{m} 
    + \kappa_2 \mathbf{u} 
    \,,
\end{equation}
with $\kappa_1 = \mathbf{n}^T\! \bm{\phi}$ 
and  $\kappa_2 = \mathbf{v}^T\! \bm{\phi}$.
Similar to the rank-one case, we can apply the recursive insertion of the 
fixed point and partial integration, \cref{eq:kappa_partial,eq:insert_x}, 
to compute the two-component vector $\bm{\kappa} = (\kappa_1, \kappa_2)$. 
We arrive at
\begin{equation}
    \label{eq:kappa_ev}
    Q_\lambda \bm{\kappa}
    =
    \frac{1}{\langle\phi'\rangle}
    \bm{\kappa} \,.
\end{equation}
This equation has two consequences: 
First, we find that $\lambda = 1 / \langle \phi' \rangle$, 
because both sides are eigenvalues of $Q_\lambda$, see \cref{eq:eigvals_r2}.
Second, the feedback vector $\bm{\kappa}$ is the corresponding eigenvector.
This gives rise to three situations: 
\begin{enumerate}[label=(\roman*)]
    \item If $Q_\lambda$ has two distinct eigenvalues, 
    one of them is equal to $\lambda$. The corresponding eigenvector
    determines the direction of $\bm{\kappa}$. 
\end{enumerate}
In the case of degeneracy, the geometric multiplicity, 
that is, the number of eigenvectors, determines the situation. 
\begin{enumerate}[label=(\roman*)]
\setcounter{enumi}{1}
    \item 
    If there is only one eigenvector, the direction of $\bm{\kappa}$ is determined
    uniquely. 
    \item If $\lambda$ has two corresponding eigenvectors, any 
    direction is a solution. The length of 
    $\bm{\kappa}$ is determined below, \cref{eq:delta_circle},
    and we obtain a ring attractor \cite{mastrogiuseppe2018linking}.
    This situation arises in the case of precise symmetry, $Q_\lambda = \lambda \mathds{1}$.
\end{enumerate}
Finally, the length of $\bm{\kappa}$ is determined by 
the variance $\Delta^0 = \mathbf{x}^T\!\mathbf{x} / N$ of the fixed point, which obeys
\begin{equation}
    \label{eq:delta_circle}
    \Delta^0 = g^2 \langle \phi^2 \rangle + \kappa_1^2 + \kappa_2^2 \,.
\end{equation}

The fixed point stability is calculated based on the techniques introduced above;
details can be found in Appendix \ref{sub:stability_for_rank_2_perturbation}.
The result is the same as that in the rank-one case:
the stability eigenvalues obey the same equations as before. Namely, 
if the spectrum of $J + \mathbf{mn}^T +\mathbf{uv}^T$ 
has the outliers $\{\lambda_1, \dots, \lambda_K\}$, 
there are always two stability eigenvalues $\gamma_\pm$, 
both with real parts smaller than one. At a fixed point $\mathbf{x}^{(i)}$, 
there are $K - 1$ additional outliers 
$\gamma_j = \lambda_j / \lambda_i$ for $j \ne i$. 
This implies that linear dynamics around a
fixed point is completely determined by its statistics and the spectrum 
of the connectivity matrix: as long as the outliers are the same,
the stability eigenvalues are independent of the rank of the perturbation $P$
or its correlations to $J$. 

This also answers our question about whether the correlation between
fixed points is responsible 
for the strong influence on each other. 
The rank-two case, too, can be analyzed by replacing the feedback 
$\kappa_1, \kappa_2$ with two constant scalars. The corresponding 
manifold is now two-dimensional, and fixed points can be
arbitrarily uncorrelated. In \cref{fig:dynamics_2fps_uncorr_g_08}, 
we show an example: plotting the projection of the fixed points along 
the vectors $\mathbf{m}$ and $\mathbf{u}$ shows that the fixed points are 
almost orthogonal. Yet, the spectra at the origin and at each fixed point
are identical to the corresponding rank-one case 
(compare with \cref{fig:dynamics_2fps_rank1_corr_g_08}).
The correlation between fixed points is hence not important for the mutual influence of different 
fixed points.

\section{Discussion}
\label{sec:discussion}
Given a network with connectivity consisting of a random and a structured part, 
we examined the effects of correlations between the two. 
We found that such correlations enrich the functional repertoire of the network. 
This is reflected in the number of non-trivial fixed points and the spectrum of the connectivity matrix. 
We analyzed precisely which aspects of the correlations determine the fixed points and eigenvalues.

In our model, the overlaps $\theta_k = \mathbf{n}^T J^k \mathbf{n}$ quantify the correlations 
between random connectivity $J$ and structured, low-rank connectivity $\mathbf{mn}^T$.
For uncorrelated networks, only $\theta_0$ is nonzero, and the spectrum of the
joint connectivity matrix has only a single outlier \cite{rajan2006eigenvalue,tao2013outliers}.
We showed that in correlated networks with $\theta_k$ nonzero for higher $k$,
multiple outliers can exist, and that with such, multiple fixed points 
induced by a random plus rank-one connectivity structure become possible. 
The correlations between random part and rank-one structure hence enrich the 
dynamical repertoire in contrast to networks with uncorrelated rank-one structures, 
which can only induce a single fixed point \cite{mastrogiuseppe2018linking}. 
Note, however, that our assumption of Gaussian connectivity limits the resulting dynamics to a single stable fixed point (discussed below).

Apart from multiple fixed points, the correlated rank-one structure can also lead to a pair of complex conjugate outliers, which in turn yield oscillatory dynamics on a limit cycle. 
In absence of correlations, such dynamics need the perturbation to be at least of rank two 
\cite{mastrogiuseppe2018linking}.
Finally, we found that correlations amplify the perturbation due to the structured components: 
the norm of a correlated rank-one structure inducing a fixed point 
decreases with increasing variance of the
random part, pointing towards possible benefits of large initial random connectivity. 

Constraining the model to Gaussian connectivity allowed us to analytically understand the
mechanisms of correlations in a nonlinear network. 
We established a remarkable one-to-one correspondence between the outliers of the connectivity matrix and 
fixed points of the nonlinear dynamics: each real outlier larger than one induces a single fixed point.
Surprisingly, the stability of the fixed points is governed by a simple set of equations 
and also only depend on the outliers of the spectrum at the origin. 
Through these results, we were able to look at the system at one point in phase space 
(the origin)
and determine its dynamics at a different part of the phase space. 
It remains an open question to which degree these insights extend to non-Gaussian 
connectivity. 
Interesting other connectivity models might include sparse connectivity 
\cite{neri2012spectra,neri2016eigenvalue,metz2019spectral}, 
different neuron types \cite{aljadeff2015transition}, 
or networks of binary neurons such as the Hopfield model \cite{hopfield1982neural}.

Our approach allows us to gain mechanistic insight into the computations underlying 
echo state and FORCE learning models which have the same connectivity structure 
as our model \cite{jaeger2004harnessing,sussillo2009generating}.
Here, the readout vector $\mathbf{n}$ is trained, which leads to correlations to 
the random part $J$ \cite{rivkind2017local,mastrogiuseppe2019geometrical}. 
Our results on multiple fixed points and oscillations show that 
these correlations are crucial for the rich functional repertoire.
However, constraining our theory to Gaussian connectivity limits the insights, 
since the learning frameworks do not have this constraint. 
One study analyzing such non-Gaussian rank-one connectivity in the 
echo state framework shows that, like in our study, each fixed point had one corresponding outlier 
in the connectivity matrix \cite{rivkind2017local}. 
However, multiple stable fixed points were observed, which is 
in contrast to our model where the Gaussian connectivity only permits the fixed 
point with largest variance to be stable.
It would thus be interesting to extend our model beyond the Gaussian statistics. 

We pointed out a general limitation of networks with random plus rank-one connectivity: 
the restriction of fixed points to a one-dimensional manifold. 
This insight is independent of the Gaussian assumption and leads to high correlations between fixed points.
Such correlations have been found to impede sequential learning of multiple fixed points \cite{beer2019one}. 
An extension to rank-two structures allows for uncorrelated fixed points. 
Surprisingly, however, the strong influence of the largest
outliers on the stability of fixed points still exists for Gaussian rank-two 
connectivity. 
Indeed, the fixed point statistics and their stability is determined solely by 
the spectral outliers of the connectivity matrix, independently of how 
these outliers were generated.
Since these relations do not hold in the non-Gaussian case, \cite{rivkind2017local}, 
we conclude that the Gaussian assumption poses a severe limitation to the space of solutions. 

Further in accordance with the echo state and FORCE learning frameworks 
\cite{jaeger2004harnessing,sussillo2009generating}, 
we model the correlations to be induced by one of the two vectors forming the rank-one structure. 
Some of the results, such as the overlaps $\theta_k$, are symmetric under
the exchange of the two vectors and should hence be unaffected.
The result on the strongly increasing norm of the perturbation when placing 
multiple outliers, on the other hand, may depend on this assumption
\cite{logiaco2019model}.
To which degree our results or the capabilities of trained networks are limited by this constraint is not clear. 

Our choice to model the structured part as a low-rank matrix was in part motivated by the computational models discussed above. 
Besides these, the existence of such structures may also be inspired by a biological perspective. 
Any feedback loop from a high-dimensional network
through an effector with a small number of degrees of freedom may be considered
as a low-rank perturbation to the high-dimensional network.
Similarly, feedback loops from cortex through basal ganglia have been modeled
as low-rank connectivity \cite{logiaco2019model}. 
Even without such explicit loops, networks may effectively have such structure if 
their connectivity is scale-free or contains hubs \cite{rivkind2019scale}.
Finally, low-rank connectivity also appears outside of neuroscience, 
for example in an evolutionary setting \cite{furusawa2018formation}. 
Whether low-rank matrices arise in general in learning networks, and to which
degree such structure is correlated with the initially present connectivity 
are interesting future questions to be approached with the theory we developed here. 

\begin{acknowledgments}
This work was supported in part by the Israeli Science Foundation (grant number 346/16, OB).
The project was further supported by the Programme Emergences of the City of Paris, ANR project MORSE (ANR-16-CE37-0016), the program “Ecoles Universitaires de Recherche” launched by the French Government and implemented by the ANR, with the reference ANR-17-EURE-0017.
F.S. acknowledges the Max Planck Society for a Minerva Fellowship.
\end{acknowledgments}

\appendix

\section{Finite number of solutions for outliers}
\label{sub:finite_solutions}
\cref{eq:eigval_eq_series} in the main text is a polynomial of infinite degree. It may thus seem that there are infinite solutions for the outlying eigenvalues, even in the case of a finite network of size $N$. We now show why this is not the case. Note that the equation was only valid for outliers (enabling the series expansion).
Denote the eigenvalues of $J$ by $\hat{\lambda}_i$, $1 \le i \le N$. 
If we diagonalize 
$J = L^T \hat{\Lambda} R$ with 
$\hat{\Lambda} = \mathrm{diag}(\hat{\lambda_1}, \dots, \hat{\lambda}_N)$, 
and write $\hat{\mathbf{n}} = L \mathbf{n}$ and $\hat{\mathbf{m}} = R \mathbf{m}$, 
then the series becomes
\begin{equation}
\lambda 
= \sum_{k=0}^\infty \frac{\sum_{i=1}^N \hat{n}_i \hat{\lambda}_i^k \hat{m}_i}{\lambda^k} 
=\sum_{i=1}^N  \frac{\hat{n}_i \hat{m}_i}{1 - \frac{\hat{\lambda}_i}{\lambda}}  \,,
\end{equation}
which is a polynomial equation of degree $N$. 

\section{Construction of the vector \textbf{n}}%
\label{sub:construction_of_the_vector_n}
In our model, $J$ and $\mathbf{m}$ are drawn independently from Gaussian 
distributions. We construct $\mathbf{n}$ from the these two quantities
for a target set of overlaps $\hat{\theta}_k$. 
Specifically, we set
\begin{equation}
    \label{eq:construct_n}
    \mathbf{n} = 
    \frac{1}{N}\sum_{k=0}^\infty 
    \frac{\hat{\theta}_k}{g^{2k}} J^k \mathbf{m} \,.
\end{equation}
Below we show that with this definition the actual overlaps
$\theta_k$ converge to the targets
$\hat{\theta}_k$ with increasing network size $N$.
Note that the scaling by $1 / N$ renders the $\theta_k$ order one quantities.

We start by analyzing the uncorrelated case, for which $\hat{\theta}_k = 0$
for any $k \ge 1$, and 
\begin{equation}
    \mathbf{n} = \frac{\hat{\theta}_0}{N} \mathbf{m} \,.
\end{equation}
We need to show that $\theta_0$ converges to $\hat{\theta}_0$ as $N \to \infty$.
To this end, we will show that the expected value has this limit and 
that the variance vanishes. 
We calculate the scalar product
\begin{equation}
    \theta_0 = \mathbf{n}^T\!\mathbf{m} 
    = \hat{\theta}_0 \, \frac{\mathbf{m}^T\!\mathbf{m}}{N} \,.
\end{equation}
The expected value of $\mathbf{m}^T\!\mathbf{m} / N$ is one, so we observe that, 
indeed, $\mathbb{E}[\theta_0] = \hat{\theta}_0$. For the variance we have
\begin{equation}
\begin{split}
    \mathrm{var}\left(\frac{\mathbf{m}^T\!\mathbf{m}}{N}\right)
    &= \frac{1}{N^2} \sum_{ij} \mathbb{E}[m_i^2 m_j^2] - 1
 \\ &= 
    \frac{1}{N^2} 
    \sum_{ij} \mathbb{E}[m_i^2] \mathbb{E}[m_j^2] - 1
 \\ &\quad + \frac{1}{N^2} 
    \left(\sum_{i} \mathbb{E}[m_i^4] - \sum_{i} \mathbb{E}[m_i^2]^2
    \right)
\\&= \mathcal{O}\left(\frac{1}{N}\right) \,.
\end{split}
\end{equation}
The term in the second line vanishes, and the one in line three is of order 
$\mathcal{O}(1 / N)$ since the fourth moment does not depend on the network size. 
The scalar product hence has a self-averaging quality in the sense that it converges
to its expected value with deviations on the order $\mathcal{O}(1 / \sqrt{N})$.

The next overlap is treated similarly:
\begin{equation}
    \theta_1 = 
    \mathbf{n}^T\! J \mathbf{m} = 
    \frac{\hat{\theta}_0}{N} 
    \sum_{ij} m_i J_{ij} m_j \,.
\end{equation}
Here, the expected value is equal to zero because of the independence between 
$J$ and $\mathbf{m}$. The variance decays with $N$ as before:
\begin{equation}
\begin{split}
    \mathrm{var}\left(\theta_1\right)
    &= 
    \left(\frac{\hat{\theta}_0}{N} \right)^2
    \sum_{ijkl} \mathbb{E}[m_i J_{ij} m_j \, m_k J_{kl} m_l] 
 \\ &= 
    \left(\frac{\hat{\theta}_0}{N} \right)^2
    \sum_{ij} \mathbb{E}[m_i^2 J_{ij}^2 m_j^2] 
 \\ &= 
    \left(\frac{\hat{\theta}_0}{N} \right)^2
    \sum_{ij} \mathbb{E}[m_i^2 m_j^2] \,\frac{g^2}{N}
\\&= \theta_0^2 g^2 \left(\frac{1}{N} + \frac{2}{N^2}\right)
\\&= \frac{\theta_0^2 g^2}{N} + 
    \mathcal{O}\left(\frac{g^2}{N^2}\right) \,.
\end{split}
\end{equation}
The off-diagonal terms in the first line disappear since entries with different 
indices are independent. Similar calculations can be done for any of the higher
order overlaps $\theta_k$, and one finds that the variance of these terms is
\begin{equation}
    \mathrm{var}(\theta_k) 
    = \frac{\theta_0^2 g^{2k}}{N} + \mathcal{O}\left(\frac{g^{2k}}{N^2}\right) \,.
\end{equation}

We now turn to the correlated case. Here, the terms $\hat{\theta}_k$ may be 
non-zero. We only discuss the simplest case with $\hat{\theta}_k = 0$ for
any $k \ge 2$, since all other cases can be treated similarly.
We write
\begin{equation}
    \mathbf{n} = 
    \frac{1}{N} \left(
    \hat{\theta}_0 \mathbf{m} +
    \frac{\hat{\theta}_1}{g^2} J \mathbf{m} 
\right) \,,
\end{equation}
and calculate the zeroth overlap
\begin{equation}
    \theta_0 
    = \hat{\theta}_0 \, \frac{\mathbf{m}^T\!\mathbf{m}}{N} 
    + 
    \frac{\hat{\theta}_1}{g^2}\, 
    \cancel{
    \frac{\mathbf{m}^T\! J \mathbf{m}}{N}}
    \,.
\end{equation}
The crossed-out term self-averages to zero due to the independence between 
$J$ and $\mathbf{m}$, so that $\mathbb{E}[\theta_0] = \hat{\theta_0}$.

Similar reasoning applies to the first overlap:
\begin{equation}
    \theta_1
    = \hat{\theta}_1 \, 
    \cancel{\frac{\mathbf{m}^T\! J^T \mathbf{m}}{N} }
    +
    \frac{\hat{\theta}_1}{g^2}\, 
    \frac{\mathbf{m}^T\! J^T J \mathbf{m}}{N}
    \,.
\end{equation}
Now, however, it is the first term that vanishes. The second one remains order $\mathcal{O}(1)$:
\begin{equation}
\begin{split}
    \mathbb{E}\left[
    \frac{\mathbf{m}^T\! J^T J \mathbf{m}}{N}
    \right]
    &= 
    \frac{1}{N}
    \sum_{ijk} \mathbb{E}[m_i J_{ji} J_{jk} m_k]
    \\ &= 
    \frac{1}{N}
    \sum_{ij} \mathbb{E}[m_i^2] \mathbb{E}[J_{ji}^2]
 \\ &= g^2 \,,
\end{split}
\end{equation}
and hence $\mathbb{E}[\theta_1] = \hat{\theta}_1$.
Similar calculations show that 
\begin{equation}
    \mathrm{var}\left(\frac{\mathbf{m}^T\! J^T J \mathbf{m}}{N}\right)
    = \frac{2 g^{4}}{N} + \mathcal{O}\left(\frac{g^4}{N^2}\right) \,,
\end{equation}
where the factor 2 is of combinatorial nature, counting how many index 
combinations yield nonzero expectations. 

The last calculations explain the scaling by $1 / g^{2k}$
in the definition of $\mathbf{n}$, \cref{eq:construct_n}.
It also points to a general 
feature of the algebra of scalar products: products of the sort 
$\mathbf{a} (J^T)^{k} J^l \mathbf{b}$ for vectors $\mathbf{a}$ and $\mathbf{b}$ 
independent of $J$
yield the expected value
\begin{equation}
    \label{eq:vector_algebra}
    \mathbb{E}\left[\mathbf{a}^T\! (J^T)^{k} J^l \mathbf{b} \right]
=
\begin{cases} g^{2k} 
    \mathbf{a}^T\! \mathbf{b} 
    & \mathrm{if} \quad k = l \,,\\
    0 &\mathrm{else.}
\end{cases} 
\end{equation}
Applying this algebra, we see that for the construction of $\mathbf{n}$
according to \cref{eq:construct_n} one indeed obtains
$\mathbb{E}[\theta_k] = \hat{\theta}_k$, valid in expectation and with variances on the 
order $\mathcal{O}(C_k g^{2k} / N)$, with a combinatorial factor $C_k$.

\section{Construction of outliers}
\label{sub:construction_of_outliers}
Here we detail how to construct a rank-one perturbation $\mathbf{mn}^T$
such that the joint matrix $J + \mathbf{mn}^T$ has a set 
$\Lambda = \{\lambda_1, \dots, \lambda_K\}$ of $K$ outliers. 
Applying \cref{eq:construct_n} for $\mathbf{n}$, the question
reduces to finding the coefficients $\hat{\theta}_k$. 

The procedure of determining the $\hat{\theta}_k$ from the $\lambda_i$ 
allows some choices. We start by choosing whether the 
$\hat{\theta}_k$ should form a truncated series or decay exponentially.
For the truncated case with $\hat{\theta}_k = 0$ for all $k \ge K$, 
the equation follows directly from outlier equation \eqref{eq:eigval_eq_series}:
\begin{equation}
    \label{eq:lambda_poly_trunc}
    0 = \lambda^{K} - \sum_{k = 0}^{K-1} \hat{\theta}_k \lambda^{K-1-k} \,.
\end{equation}
To obtain $K$ outliers from a non-truncated series, one can write 
the $\hat{\theta}_k$ as sums of exponentially decaying terms
\begin{equation}
    \label{eq:theta_k_exp}
    \hat{\theta}_k = \sum_{\alpha = 1}^K a_\alpha b_\alpha^k  \,,
\end{equation}
with coefficients $a_\alpha$ and bases $b_\alpha$. 
Evaluating the geometric series leads to the polynomial equation 
\begin{equation}
    \label{eq:lambda_poly_exp}
    1 = \sum_{\alpha = 1}^K \frac{a_\alpha }{\lambda - b_\alpha} \,.
\end{equation}
Either choice hence yields a polynomial of degree $K$, the roots
of which need to be the target outliers $\lambda_i$.
The coefficients of this polynomial can hence be obtained from
a comparison with the coefficients of the polynomial 
$p(\lambda) = \prod_{i=1}^K (\lambda - \lambda_i)$.
For example, in the case of a truncated series, the coefficients $\hat{\theta}_k$ 
are determined by 
\begin{equation}
    \hat{\theta}_k = (-1)^k \sum_{\pi \in S_{\binom{K}{k+1}}} 
    \left(
    \prod_{i \in \pi} \lambda_i \right)
        \,,
\end{equation}
where $S_{\binom{K}{k}}$ denotes the possible choices of $k$ different indices from the 
$K$ available ones. Note that the indices for $\theta_k$ run from zero to $K-1$, whereas
those of $\lambda_i$ run from 1 to $K$.

In the case of exponentially decaying series, one can derive a similar equation 
from \cref{eq:lambda_poly_exp}. However, there are $2K$ parameters
$a_\alpha$ and $b_\alpha$ -- there is the freedom to choose $K$ of them, as long 
as $b_\alpha < g$.

\section{Least square vector \textbf{n}}%
\label{sub:least_square_vector_n}
Here we introduce another method of constructing multiple outliers.
Assume that $J$, $\mathbf{m}$ are given and we want to find the least square vector 
$\mathbf{n}$ such that we have the set of outliers $\{\lambda_1, \dots \lambda_K\}$.
We apply the matrix determinant lemma, 
\cref{eq:matrix_det_lemma}, and obtain
\begin{equation}
    \mathbf{n}^T (\lambda_\alpha \mathds{1} - J)^{-1} \mathbf{m} = 1 
    \quad \forall \quad \alpha \in \{1, \dots, K\} \,.
\end{equation}
This can be read as an underconstrained linear system 
$A \mathbf{n} = \mathbf{1}$,
with the vector of ones $\mathbf{1}$ and the
matrix $A \in \mathbb{C}^{K \times N}$ defined by its rows
\begin{equation}
    A^T_{\alpha} = \frac{1}{\lambda_\alpha} \mathbf{w}_{\lambda_\alpha}
    \,,
\end{equation}
with 
$\mathbf{w}_\lambda = \left(\mathds{1} - \frac{J}{\lambda}\right)^{-1} \mathbf{m}$.
Since $K < N$, the matrix $A$ is singular. The least square solution to the 
system is given by 
\begin{equation}
    \label{eq:nA1}
    \mathbf{n} = A^{+} \mathbf{1} \,,
\end{equation}
with the pseudoinverse of $A$ 
denoted by $A^{+} = A^T (AA^T)^{-1} \in \mathbb{C}^{N \times K}$.

We express the vector 
$\mathbf{w}_\lambda$ by the series expansion
\begin{equation}
    \mathbf{w}_{\lambda} =
    \sum_k \left(\frac{J}{\lambda}\right)^k \mathbf{m}  \,,
\end{equation}
and insert this into the term $AA^T$. Applying the
algebra \eqref{eq:vector_algebra} developed above yields
\begin{equation}
    \label{eq:AAT}
    \begin{split}
    (AA^T)_{\alpha\beta} 
    &= \sum_{i=1}^N A_{\alpha i} A_{\beta i} \\
    &= \frac{1}{\lambda_\alpha \lambda_\beta} \sum_{k, l} 
    \mathbf{m}^T\! \left(\frac{J^T}{\lambda_\alpha}\right)^k
    \left(\frac{J}{\lambda_\beta}\right)^l \mathbf{m} \\
    &= \frac{1}{\lambda_\alpha \lambda_\beta} \sum_{k} 
    \left(\frac{g^2}{\lambda_\alpha\lambda_\beta}\right)^k \mathbf{m}^T\! \mathbf{m} \\
    &= \frac{N}{\lambda_\alpha \lambda_\beta - g^2} \,.
    \end{split} 
\end{equation}
Inserting this into \cref{eq:nA1}, we can finally write
\begin{equation}
    \mathbf{n} 
    = \frac{1}{N}\sum_{\alpha=1}^K a_\alpha \mathbf{w}_{\lambda_\alpha} \,,
\end{equation}
where the coefficients $a_\alpha$ are determined by solving the linear equation
\begin{equation}
    1 = \sum_{\beta = 1}^K
        \frac{a_\beta}{\lambda_\alpha - g^2 / \lambda_\beta}
    \quad \forall \quad \alpha \in \{1, \dots, K\} \,.
\end{equation}
Connecting to the above schemes of constructing $\mathbf{n}$ with deliberate overlaps
$\theta_k = \mathbf{n}^T\! J^k \mathbf{m}$, we compute these for the least 
square solution found here. We find that overlaps decay exponentially as in 
\cref{eq:theta_k_exp}, namely
\begin{equation}
    \theta_k = \sum_{\alpha = 1}^{K} a_\alpha \left(\frac{g^2}{\lambda_\alpha}\right)^k  \,.
\end{equation}

We observed numerically that for a large number of outliers $K$ the 
rank-one perturbation became the dominant term in the matrix $J + \mathbf{mn}^T$. 
To understand this, we look at its Frobenius norm, given by 
\begin{equation}
    ||\mathbf{mn}^T||^2 
    = \mathrm{Tr}((\mathbf{mn}^T)^T \mathbf{mn}^T)
    = N \mathbf{n}^T\!\mathbf{n} \,.
\end{equation}
The squared norm of $\mathbf{n}$ can be obtained from the pseudoinverse defined above:
\begin{equation}
    \mathbf{n}^T\!\mathbf{n} 
    = \mathbf{1}^T (A^+)^T A^+ \mathbf{1} 
    = \mathbf{1}^T (A A^T)^{-1} \mathbf{1}  \,.
\end{equation}
To arrive at a general expression for this quantity, we calculated
\cref{eq:AAT} explicitly for the cases $K = 1$ and $K = 2$ and performed thorough
numerical checks for larger $K$ (see 
\cref{fig:scaling_with_outliers_N_1000_g_08}). The resulting 
equation is 
\begin{equation}
    \mathbf{n}^T\!\mathbf{n} 
    = \frac{g^2}{N} \left(\prod_{\alpha=1}^K \frac{\lambda_\alpha^2}{g^2} - 1\right)  \,.
\end{equation}
For an increasing number of outliers, the offset by minus one
becomes negligible, and we arrive at the result stated in the main text, \cref{eq:scaling_mn}.

\section{Mean field theory with correlations}%
\label{sub:mean_field_theory_with_correlations}
We analyze fixed points of the form 
\begin{equation}
    \label{eq:x_fp_I}
    \mathbf{x} = J \bm{\phi} + \kappa \mathbf{m} + \mathbf{I} \,.
\end{equation}
We extend the setting in the main text 
by allowing for a constant input vector $\mathbf{I}$. 
Like the other vectors, we assume $\mathbf{I}$ to be Gaussian and uncorrelated 
to $J$.

We want to compute $\kappa = \mathbf{n}^T\!\phi(\mathbf{x})$. The mean field assumption 
is that $\mathbf{x}$ is a Gaussian variable. That is, its entries $x_i$ are drawn 
from a normal distribution with zero mean and variance $\Delta^0$. The 
variance $\Delta^0$ is determined self-consistently below. Since the entries of the 
vector follow the same statistics, we look at a representative $x_i$ and
drop the index $i$:
\begin{equation}
    \label{eq:x_single}
    x = \sqrt{\Delta^0} z_x \,,
\end{equation}
with the standard Gaussian random variable $z_x \sim \mathcal{N}(0, 1)$.
By the model assumption, the vector $\mathbf{n}$ is also a Gaussian. 
One can express $\mathbf{n}$ explicitly in relation to $\mathbf{x}$ by defining 
\begin{equation}
    \label{eq:nx}
    n = \frac{1}{N} \left(\sigma_n \sqrt{1 - \rho} \,z_n 
        + \sigma_n \rho \, z_x \right) \,,
\end{equation}
with a second, independent standard Gaussian random variable $z_n$. The parameters 
$\sigma_n^2$ and $\rho$ encode the variance of $\mathbf{n}$ and its correlation to 
$\mathbf{x}$. The self-averaging quality of the scalar product allows us to write
\begin{equation}
    \mathbf{n}^T\!\mathbf{x} = N \,\mathbb{E}[n x]
    = \sqrt{\Delta^0} \sigma_n \rho \,.
\end{equation}

Computing $\kappa$ can then be achieved by exchanging between the scalar product 
$\mathbf{n}^T\!\bm{\phi}$ and the corresponding Gaussian integral:
\begin{equation}
    \label{eq:kappa_derive_explicit}
    \begin{split}
    \kappa 
    &= \mathbf{n}^T\! \bm{\phi} \\
    &= \int \!\mathcal{D}z_x \int \!\mathcal{D}z_n
        \left(\sigma_n \sqrt{1 - \rho} \,z_n 
        + \sigma_n \rho \, z_x \right) 
        \phi(\sqrt{\Delta^0} \, z_x) \\
    &= \sqrt{\Delta^0} \sigma_n \rho 
        \int \!\mathcal{D}z_x \phi'(\sqrt{\Delta^0} \, z_x) \\
    &= \mathbf{n}^T\! \mathbf{x} \, \langle \phi' \rangle \,,
    \end{split}
\end{equation}
where $\mathcal{D}z$ is the standard Gaussian measure. 
In the second line, the term $\sigma_n\sqrt{1 - \rho} z_n$ vanishes with the integration 
over $z_n$. For the second summand, the integrating over $z_n$ evaluates to 1. 
The step from second to third line involves partial integration: for some function $f$ and the 
Gaussian variable $z \sim \mathcal{N}(0, 1)$, we have 
$\int \mathcal{D}y \,z f(z)  = \int \mathcal{D}z\, f'(z) $.
This is also known as Stein's lemma \cite{mastrogiuseppe2018linking}.
Finally, the angled brackets $\langle\cdot\rangle$ indicate the average over the fixed 
point statistics, 
\begin{equation}
    \label{eq:dphi_avg_meth}
    \langle \phi' \rangle
    = \int \mathcal{D}z \,\phi'(\sqrt{\Delta^0} z)  \,.
\end{equation}

Such explicit representations of pairs of correlated Gaussian variables 
have been applied before 
\cite{rivkind2017local,mastrogiuseppe2018linking}. 
Below, however, we encounter 
a multitude of such Gaussian vectors, so such a framework would become increasingly
cumbersome. We thus introduce a new formalism which allows us to model arbitrary many 
Gaussian vectors. 
Let $\mathbf{a}$ and $\mathbf{b}$ be two such vectors of interest. Like before, we are 
interested in the statistics of a representative entry, namely $a$ and $b$. 
We define these as
\begin{equation}
    a 
    = B_a \cdot X
    \,, 
    \qquad 
    b 
    = B_b \cdot X \,.
\end{equation}
Here, $B_a, B_b \in \mathbb{R}^K$ are a set of real-valued coefficients 
and $X$ is a $K$-dimensional standard normal Gaussian variable 
with mutually independent entries  $X_\alpha \sim \mathcal{N}(0, 1)$.
The $K$-dimensional dot product is defined as 
$B_a \cdot X 
    = \sum_{\alpha = 1}^K (B_a)_\alpha X_\alpha$.
Any additional vectors $\mathbf{c, d,} \dots$
are added by defining corresponding coefficients $B_c, B_d, \dots$. 
One just needs to choose the dimension $K$ of the embedding to be sufficiently
large. 

Since scalar products in the $N$-dimensional physical space 
are self-averaging, we can write:
\begin{equation}
    \label{eq:define_coeffs}
    \frac{1}{N} \mathbf{a}^T\!\mathbf{b}  
    = \frac{1}{N}\sum_{i=1}^N a_i b_i =  B_a \cdot B_b  \,.
\end{equation}
In line with the previous sections, the equality sign is only valid in the
limit $N \to \infty$, and the variance decays like $1 / N$.
Ultimately, we are interested in such scalar products. This allows us 
to use the coefficients $B_a$ merely as placeholders inside calculations, 
without ever defining their actual values. 

With this notation we return to the computation of $\kappa$:
\begin{equation}
    \label{eq:kappa_derive}
    \begin{split}
    \kappa 
    &= \mathbf{n}^T\! \bm{\phi} \\
    &= N\int \mathcal{D}X \, (B_n \cdot X) \, \phi(B_x \cdot X) \\ 
    &= N B_n \cdot B_x \, \int \mathcal{D}X \,\phi'(B_x \cdot X) \\ 
    &= \mathbf{n}^T\! \mathbf{x} \, \langle \phi' \rangle \,,
    \end{split}
\end{equation}
where $\mathcal{D}X$ is the now standard Gaussian measure in $K$ dimensions. 
The third line is obtained using partial integration as before. 
In the last line of \cref{eq:kappa_derive} above, 
we inserted the definition of coefficients from above, 
\cref{eq:define_coeffs}, for the scalar products. 

The advantage of the new formalism is that it allows us to continue the
calculation. We insert the fixed point 
$\mathbf{x} = J \bm{\phi} + \kappa \mathbf{m} + \mathbf{I}$.
In the case of structure vectors drawn independently from $J$, the term
$\mathbf{n}^T\! J \bm{\phi}$ vanishes and one recovers the known result
$\kappa = \langle\phi'\rangle \,\mathbf{n}^T\!\left(\kappa \mathbf{m} + \mathbf{I}\right)$
\cite{mastrogiuseppe2018linking}.
For the general case,
we go on calculating $\mathbf{n}^T\! J \bm{\phi}$.
The scalar product allows to pull the random matrix to the left side, 
and hence
\begin{equation}
    \label{eq:kappa2_derive}
    \begin{split}
    \mathbf{n}^T\! J \bm{\phi} 
    &= N\int\, \mathcal{D}X (B_{J^T\!n} \cdot X) \, \phi(B_x \cdot X) \\ 
    &= \langle \phi' \rangle \, \mathbf{n}^T\! J \mathbf{x} \,.
    \end{split}
\end{equation}
Recursively applying this strategy, we arrive at
\begin{equation}
    \label{eq:kappa1}
\begin{split}
    \kappa
    &= 
    \langle\phi'\rangle \,
    \mathbf{n}^T
    \sum_k 
    \left(\langle\phi'\rangle J \right)^k
    (\kappa\mathbf{m} + \mathbf{I})  
    \\ &=
    \langle\phi'\rangle \,
    \mathbf{n}^T\! M (\kappa\mathbf{m} + \mathbf{I})  
    \,,
\end{split}
\end{equation}
with
\begin{equation}
    M
    = \left(\mathds{1} - \langle \phi' \rangle J \right)^{-1} \,.
\end{equation}

For a driven network with nonzero $\mathbf{n}^T\!M\mathbf{I}$, one can compute 
$\kappa$ by re-sorting:
\begin{equation}
    \label{eq:kappa}
    \kappa = 
    \frac{ \langle \phi' \rangle \mathbf{n}^T\! M \mathbf{I}}
        {1 -  \langle \phi' \rangle\mathbf{n}^T\! M \mathbf{m}} 
    \,.
\end{equation}
The scalar $\langle\phi'\rangle$, \cref{eq:dphi}, is a function of the 
fixed point variance $\Delta^0 = \mathbf{x}^T\mathbf{x} / N$.
Due to the independence of $\mathbf{m}$ and $\mathbf{I}$ from 
$J$, the variance obeys the same equation as in previous studies
\cite{mastrogiuseppe2018linking}:
\begin{equation}
    \label{eq:var_I}
    \Delta^0 
    = g^2 \langle \phi^2 \rangle 
    + \left(\kappa \mathbf{m} + \mathbf{I}\right)^T\! \left(\kappa \mathbf{m} + \mathbf{I}\right) / N
    \,,
\end{equation}
with $
    \langle \phi^2 \rangle
    = \int \mathcal{D}z \,\phi^2(\sqrt{\Delta^0} z) 
$.
The coupled nonlinear 
\cref{eq:kappa,eq:dphi,eq:var_I}
can be solved numerically if the overlaps $\theta_k = \mathbf{n}^T\!J^k\mathbf{m}$ 
and $\mathbf{n}^T\!J^k\mathbf{I}$, which respectively enter 
$\mathbf{n}^T\! M \mathbf{m}$ and
$\mathbf{n}^T\! M \mathbf{I}$, are known.

In the case of no input, $\mathbf{I} = 0$, $\kappa$ is not directly determined. 
Instead, $\langle\phi'\rangle$ is given directly by the outliers via 
\cref{eq:eigval_dphi}, as discussed in the main text.
The mean field equations can be closed by numerically solving 
\cref{eq:dphi_avg_meth} for $\Delta^0$.
The corresponding $\kappa$ is determined up to the sign by 
\begin{equation}
    \label{eq:kappa_delta}
    \Delta^0 
    = g^2 \langle \phi^2 \rangle 
    + \kappa^2 
    \,.
\end{equation}

\section{Stability eigenvalues}%
\label{sub:stability_eigenvalues}
The matrix $J R' - \mathds{1} \gamma$ 
with the diagonal matrix $R_{ij}' = \delta_{ij} \phi_i'$
is invertible as long as $\mathbf{\gamma} \in \mathbb{C}$ 
is not within the spectrum of $J R'$. We apply the matrix determinant 
lemma to compute the characteristic polynomial,
\begin{equation}
\begin{split}
     &\det \left(J R' - \mathds{1} \gamma \right)  \\
    &\qquad= 
    \left(1 
    + \mathbf{n}^T\! R' (J R' - \mathds{1} \gamma)^{-1} \mathbf{m}\right) \det(J R' - \mathds{1}\gamma) 
    \,.
\end{split}
\end{equation}
The first bracket has to vanish, so that we arrive at
\begin{equation}
    \label{eq:eigval_eq_gamma_meth}
\gamma = 
\mathbf{n}^T\! R' \left(\mathds{1} - \frac{J R'}{\gamma}\right)^{-1} \mathbf{m} 
=
\sum_k \frac{\mathbf{n}^T\! R' (J R')^k \mathbf{m}}{\gamma^k}
\,,
\end{equation}
for a nonzero $\gamma$.

We calculate the terms $\mathbf{n}^T\!R' \left(JR'\right)^k \mathbf{m}$ 
applying the Gaussian mean field theory introduced above. 
For brevity we use induction. The hypothesis to be proven is
\begin{equation}
    \small
\begin{split}
    \label{eq:induct_hypo}
    &\mathbf{n}^T\!R' \left(JR'\right)^k \mathbf{m} \\
    &\,\,= 
    \mathbf{n}^T\!J^k \mathbf{m} \langle \phi' \rangle^{k+1}
    + 
    \frac{\mathbf{x}^T\!\mathbf{m}}{N} \langle \phi''' \rangle
    \sum_{l = 0}^{k} q^l 
    \mathbf{n}^T\!(\langle \phi' \rangle J)^{k-l} \mathbf{x} \,,
\end{split}
\end{equation}
where $\mathbf{x}$ denotes the fixed point and  we define
\begin{equation}
    \label{eq:def_q}
    q =g^2 \langle \phi'' \phi + \phi'^2 \rangle \,.
\end{equation}
We start the induction by calculating the Gaussian integral
\begin{equation}
    \label{eq:nTRdm}
    \begin{split}
    \mathbf{n}^T\!R' \mathbf{m} 
    &= N\int \mathcal{D}X \,(B_n \cdot X) \, \phi'(B_x \cdot X) \,B_m \cdot X\\ 
    &= \mathbf{n}^T\! \mathbf{m} \,\langle \phi' \rangle
    + 
    \mathbf{n}^T\! \mathbf{x} \,
    \frac{\mathbf{x}^T\!\mathbf{m}}{N} \,\langle \phi''' \rangle \,.
    \end{split}
\end{equation}
We used partial differentiation twice, which yields the third derivative 
of the nonlinearity. The steps above did not depend the specific 
vectors $\mathbf{m}$ and $\mathbf{n}$ so we will apply the same step below without 
explicitly mentioning the Gaussian integrals. 
The entire deviation furthermore does not depend on the vector $\mathbf{n}$. 
We make use of this independence in the induction step where we assume
the hypothesis \eqref{eq:induct_hypo} to be true also after replacing 
$\mathbf{n}^T$ with $\mathbf{n}^T\!J$. This term is identified by square brackets 
in the below calculation:
\begin{equation}
\begin{split}
    &\mathbf{n}^T\!R'\left(JR'\right)^k \mathbf{m} 
    \\&\quad=
    \mathbf{n}^T\! 
    \left( \langle \phi' \rangle + \frac{\mathbf{x}\mathbf{x}^T}{N} \langle\phi'''\rangle\right)
    \left(J R'\right)^{k} \mathbf{m}  
    \\&\quad=
    \left[\mathbf{n}^T\!J\right] R' \left(J R'\right)^{k-1} \mathbf{m} \,\langle \phi' \rangle
    \\&\qquad +
    \mathbf{n}^T\! \mathbf{x} \frac{1}{N} \mathbf{x}^T\! \left(J  R'\right)^k
    \mathbf{m} \,\langle \phi''' \rangle  
    \\&\quad=
    \left[\mathbf{n}^T\!J\right] J^{k-1} \mathbf{m}\, 
    \langle \phi' \rangle^{k}
    \langle \phi' \rangle
    \\&\qquad 
    +
    \frac{\mathbf{x}^T\!\mathbf{m}}{N} \langle \phi''' \rangle
    \sum_{l = 0}^{k-1} q^l 
    \left[\mathbf{n}^T\!J\right] 
    (\langle \phi' \rangle J)^{k-1-l} \mathbf{x} \langle\phi'\rangle
    \\&\qquad 
    +
    \mathbf{n}^T\! \mathbf{x} \,
    \frac{\mathbf{x}^T\!\mathbf{m}}{N} \,q^k \,\langle \phi''' \rangle \,.
\end{split}
\end{equation}
The first step is to replace $\mathbf{m}$ with $\left(JR'\right)^k \mathbf{m}$
in \cref{eq:nTRdm}.
The step from second to third line involves the induction hypothesis for $k - 1$, 
and including the last term in the sum completes the induction.
The last term needs to be calculated separately. We show that
\begin{equation}
    \mathbf{x}^T\! \left(J  R'\right)^k \mathbf{m} 
    = \mathbf{x}^T\!\mathbf{m} \,q^k
\end{equation}
in a separate induction. The start is trivial. For the induction step at $k \ge 1$, 
we insert the fixed point equation $\mathbf{x} = J \bm{\phi} + \kappa \mathbf{m} + \mathbf{I}$.
\begin{equation}
    \begin{split}
    \mathbf{x}^T\! \left(J  R'\right)^k \mathbf{m} 
    &=
    \bm{\phi}^T\! J^T\! \left(J  R'\right)^k \mathbf{m} 
    + \cancel{\left(\kappa \mathbf{m} + \mathbf{I}\right)^T\! \left(J  R'\right)^k \mathbf{m} } \\
    &=
    \bm{\phi}^T\! J^T\!\! J \left[R' \left(J  R'\right)^{k-1} \mathbf{m}\right]  \\
    &=
    g^2 \bm{\phi}^T\! \left[ R' \left(J  R'\right)^{k-1} \mathbf{m}\right]  \\
    &=
    q \,\mathbf{x}^T\! \left(J  R'\right)^{k-1} \mathbf{m} \,.
    \end{split} 
\end{equation}
Inserting the induction hypothesis for $k - 1$ proves the statement. A few comments 
on the steps of the calculation:
\begin{itemize}
    \item The crossed-out term in the first line vanishes because $\mathbf{m}$ and $\mathbf{I}$
        are drawn independently of $J$. Showing this formally is a matter of applying 
        the same techniques as introduced above recursively.
    \item The step from line two to three involves the vector algebra \eqref{eq:vector_algebra}
        introduced above, according to which 
        $\mathbf{a}^T\! J^T\!\!J \mathbf{b} = g^2 \mathbf{a}^T\! \mathbf{b}$ 
        for two vectors $\mathbf{a, b}$ independent of $J$.
    \item The fourth line is obtained by applying partial integration.
        For an arbitrary Gaussian vector $\mathbf{a}$, we have
        \begin{equation}
        \begin{split}
            \bm{\phi}^T\! R' \mathbf{a}
            &= N\int \!\mathcal{D}X \,\phi(B_x \cdot X) \, \phi'(B_x \cdot X) \,B_a \cdot X\\ 
            &= \mathbf{x}^T\! \mathbf{a} \langle\phi'' \phi + \phi'^2\rangle\,.
        \end{split}
        \end{equation}
\end{itemize}

We go back to the eigenvalue equation 
\eqref{eq:eigval_eq_gamma_meth}
and insert \cref{eq:induct_hypo}, 
which yields
\begin{equation}
\begin{split}
\gamma &= 
\sum_k \frac{\mathbf{n}^T\! R' (J R')^k \mathbf{m}}{\gamma^k}
\\&=
\langle\phi'\rangle
\sum_k
\left(\frac{\langle \phi' \rangle}{\gamma}\right)^k
    \mathbf{n}^T\!J^k \mathbf{m}
\\&\quad
    + 
    \frac{\mathbf{x}^T\!\mathbf{m}}{N}\, \langle \phi''' \rangle
    \sum_k \left(\frac{q}{\gamma}\right)^k
    \sum_{l = 0}^{k} \,
    \left(\frac{\langle \phi' \rangle}{q}\right)^l
    \mathbf{n}^T\!J^{l} \mathbf{x}  \,.
\end{split}
\end{equation}
Note that we swap the order of summation in the second sum. This sum evaluates to
\begin{equation}
    \small
\begin{split}
    &\sum_k \left(\frac{q}{\gamma}\right)^k \,
    \sum_{l = 0}^{k} \,
    \left(\frac{\langle \phi' \rangle}{q} J \right)^l
    \\&\quad=
    \sum_k \left(\frac{q}{\gamma}\right)^k
    \left(\mathds{1} - \frac{\langle \phi' \rangle}{q} J \right)^{-1}
    \left[
    \mathds{1} - 
    \left(\frac{\langle \phi' \rangle}{q} J \right)^{k + 1}
    \right]
    \\&\quad=
    \left(\mathds{1} - \frac{\langle \phi' \rangle}{q} J \right)^{-1}
    \left[
    \frac{1}{1 - \frac{q}{\gamma}}
    -
    \frac{\langle \phi' \rangle}{q} J 
    \left(\mathds{1} - \frac{\langle \phi' \rangle}{\gamma} J \right)^{-1}
    \right]
    \\&\quad=
    \left(\mathds{1} - \frac{\langle \phi' \rangle}{q} J \right)^{-1}
    \left[
    \frac{1}{1 - \frac{q}{\gamma}}
    \left(\mathds{1} - \frac{\langle \phi' \rangle}{\gamma} J \right)^{-1}
    \left(\mathds{1} - \frac{\langle \phi' \rangle}{q} J \right)
    \right]
    \\&\quad=
    \frac{1}{1 - \frac{q}{\gamma}}
    \left(\mathds{1} - \frac{\langle \phi' \rangle}{\gamma} J \right)^{-1} \,.
\end{split}
\end{equation}
Inserting this into \cref{eq:eigval_eq_gamma_meth} for $\gamma$, 
we obtain
\begin{equation}
    \label{eq:gamma_step}
    \gamma =
    \mathbf{n}^T
    \left(\mathds{1} - \frac{\langle\phi'\rangle}{\gamma} J \right)^{-1}
    \left[
        \langle \phi' \rangle \mathbf{m}
        + 
        \frac{\langle \phi''' \rangle}{1 - \frac{q}{\gamma}}
        \frac{\mathbf{m}^T\! \mathbf{x}}{N} \,\mathbf{x}
    \right] \,.
\end{equation}
One can further simplify this expression by inserting the fixed point
$\mathbf{x} = J \phi(\mathbf{x}) + \kappa \mathbf{m} + \mathbf{I}$ and 
evaluating the Gaussian statistics. 
In particular, we have 
\begin{equation}
    \mathbf{a}^T\!\mathbf{x} = 
    \mathbf{a}^T\!
    \left(\mathds{1} - \langle\phi'\rangle J\right)^{-1} 
    \left(\kappa \mathbf{m} + \mathbf{I}\right)
\end{equation}
for any Gaussian vector $\mathbf{a}$. 
In particular, 
\begin{equation}
\label{eq:nTx_double}
\begin{split}
    &\mathbf{n}^T
    \left(\mathds{1} - \frac{\langle\phi'\rangle}{\gamma} J \right)^{-1} \mathbf{x}
    \\&\quad=
    \mathbf{n}^T
    \left(\mathds{1} - \frac{\langle\phi'\rangle}{\gamma} J \right)^{-1}
    \left(\mathds{1} - \langle\phi'\rangle J\right)^{-1} 
    \left(\kappa \mathbf{m} + \mathbf{I}\right) \,.
\end{split}
\end{equation}
The product of the two inverses can be conveniently split. For any matrix $A$ 
and a scalar $a$, completion of the denominator yields
\begin{equation}
    \begin{split}
    &\frac{1}{1 - a} \left[
    \left(\mathds{1} - A \right)^{-1} 
    - a
    \left(\mathds{1} - a A \right)^{-1} 
    \right] 
    \\&\quad=
    \frac{1}{1 - a} \left[
    \left(\mathds{1} - aA \right)
    - a
    \left(\mathds{1} - A \right)
    \right] 
    \\&\qquad\quad \times
    \left(\mathds{1} - a A \right)^{-1}
    \left(\mathds{1} - A \right)^{-1} 
    \\&\quad=
    \left(\mathds{1} - a A \right)^{-1}
    \left(\mathds{1} - A \right)^{-1}  \,.
    \end{split}
\end{equation}
Inserting this into \cref{eq:nTx_double}, we arrive at
\begin{equation}
\begin{split}
    \gamma &=
        \langle \phi' \rangle 
    \mathbf{n}^T
    \left(\mathds{1} - \frac{\langle\phi'\rangle}{\gamma} J \right)^{-1}
    \mathbf{m}
    \\&\quad
        + 
        \frac{\langle \phi''' \rangle 
            }{\left(1 - \frac{q}{\gamma}\right)\left(1 - \frac{1}{\gamma}\right)}
            \frac{ \mathbf{m}^T\!\mathbf{x} }{N}
    \\&\qquad \times
    \left[
        \frac{\kappa}{\langle\phi'\rangle}
        -
        \frac{1}{\gamma}
    \mathbf{n}^T
    \left(\mathds{1} - \frac{\langle\phi'\rangle}{\gamma} J \right)^{-1}
    (\kappa\mathbf{m} + \mathbf{I})
    \right] \,.
\end{split}
\end{equation}
We made use of the \cref{eq:kappa} constraining $\kappa$, 
namely
\begin{equation}
    \frac{\kappa}{\langle\phi'\rangle}
    =
    \mathbf{n}^T
    \left(\mathds{1} - \langle\phi'\rangle J \right)^{-1}
    (\kappa\mathbf{m} + \mathbf{I}) \,.
\end{equation}
Re-sorting terms finally results in 
\begin{equation}
\label{eq:gamma_with_I}
\begin{split}
    &\left[
        \frac{\gamma}{\langle\phi'\rangle} 
        - \mathbf{n}^T\! \left(\mathds{1} - \frac{\langle\phi'\rangle}{\gamma} J \right)^{-1}
        \mathbf{m}
    \right]
    \\&\quad=
    \frac{\langle\phi'''\rangle \kappa (\kappa + \mathbf{m}^T\!\mathbf{I} / N) \gamma
    }{\langle\phi'\rangle (\gamma - q)(\gamma - 1)}
    \\&\qquad \times
    \left[
        \frac{\gamma}{\langle\phi'\rangle} 
        - \mathbf{n}^T\! \left(\mathds{1} - \frac{\langle\phi'\rangle}{\gamma} J \right)^{-1}
        \left(\mathbf{m} + \frac{1}{\kappa} \mathbf{I}\right)
    \right] \,.
\end{split}
\end{equation}
Note that we also inserted
$\mathbf{m}^T\!\mathbf{x}/N = \kappa + \mathbf{m}^T\!\mathbf{I}/N$.
Without knowledge about the interaction between $\mathbf{n}$ and $\mathbf{I}$, 
we cannot further simplify since the term
$\mathbf{n}^T\! 
\left(\mathds{1} - \frac{\langle\phi'\rangle}{\gamma} J \right)^{-1}\mathbf{m}$
without $\mathbf{I}$
and the other term including it, 
$\mathbf{n}^T\! \left(\mathds{1} - \frac{\langle\phi'\rangle}{\gamma} J \right)^{-1}
\left(\kappa \mathbf{m} +  \mathbf{I}\right)$,
appear in two different parts of the theory: the first one is a property of the 
matrix, related to the outlier equation 
$ \lambda  = \mathbf{n}^T\! 
\left(\mathds{1} - \frac{1}{\lambda} J \right)^{-1}\mathbf{m}$, 
the second one determines $\kappa$.

\section{Stability eigenvalues for the autonomous network}%
\label{sub:stability_eigenvalues_for_the_autonomous_network}
\begin{figure*}[tb]
    \includegraphics[width=1.\linewidth]{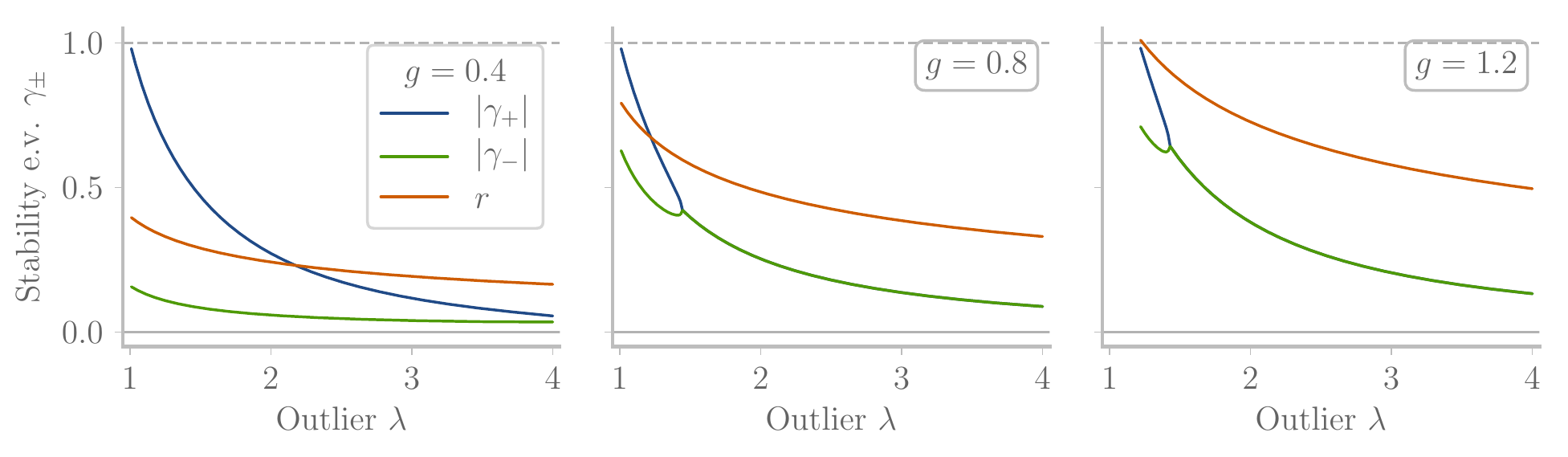}
    \caption{Stability eigenvalues $\gamma_\pm$ depend on the outlier $\lambda$
    corresponding to a fixed point. The $\gamma_\pm$ are 
    solutions to the mean field equations, fully determined by the 
    eigenvalue $\lambda$ and the random strength $g$.
    The three plots corresponds to $g \in \{0.4, 0.8, 1.2\}$. 
    Orange lines: radius of the bulk; 
    any eigenvalues with smaller magnitude will not be observable, 
    and hence not numerically testable for finite size networks. 
    Where the absolute values of $\gamma_\pm$ coincide, the two 
    form a pair of complex conjugates.
    Note that for $g > 1$, the minimal $\lambda$ to stabilize the 
    chaotic activity is larger than one.
    }
    \label{fig:stability_eigenvalues_pm}
\end{figure*}
In the autonomous case, $\mathbf{I} = \mathbf{0}$, the
square brackets in \cref{eq:gamma_with_I} become identical,
and the equation splits into a quadratic part and one of degree $K - 1$.
By applying the identity $\lambda = 1 / \langle\phi'\rangle$, 
we arrive at
\begin{equation}
    \label{eq:gamma_auto}
    0 = 
    \left[\lambda \gamma - 
    \mathbf{n}^T\!
    \left(\mathds{1} - \frac{J}{\lambda \gamma} \right)^{-1}\!
    \mathbf{m}
    \right]
    \left[
        \frac{\langle\phi'''\rangle \kappa^2 \lambda \gamma}{(\gamma - q)(\gamma - 1)} - 1
    \right] \,.
\end{equation}
The second bracket is squared in $\gamma$ and exhibits the roots 
\begin{align}
    \label{eq:gamma_pm}
    \gamma_\pm &= \frac{c}{2} \pm \sqrt{\left(\frac{c}{2}\right)^2 - q} \,,
    \shortintertext{with}
    \label{eq:def_c}
    c &= 1 + q + 
    \kappa^2
    \frac{\langle\phi'''\rangle}{\langle\phi'\rangle}
    \,, 
\end{align}
and $q =g^2 \langle \phi'' \phi + \phi'^2 \rangle$ as defined above, \cref{eq:def_q}.
We observe that $\gamma_\pm$ is entirely defined by the fixed point
statistics. One can even reduce the problem to two parameters, for example 
the outlier $\lambda$ and the network parameter $g$ which quantifies the strength of 
the random connectivity. This allows to thoroughly scan the numerical values 
of $\gamma_\pm$. The results can be observed in 
\cref{fig:stability_eigenvalues_pm}. The first observation is that 
both $\gamma_+$ and $\gamma_-$ are always smaller than one. They hence do not 
destabilize the fixed point. Additionally, we can compare $\gamma_\pm$ to the 
radius of the bulk, which is given by $r = g \sqrt{\langle \phi'^2 \rangle}$
\cite{mastrogiuseppe2018linking}.
This shows that $\gamma_-$ is always within the bulk and hence not observable numerically. 

The roots of the first bracket in \cref{eq:gamma_auto}
are identified by comparing once again with 
\cref{eq:eigval_eq} which defines the outliers $\lambda$ of 
the spectrum at the origin. In fact, the equation is identical, but now the variable is 
$\lambda \gamma$. Here, $\lambda$ is the outlier corresponding to the fixed point 
under consideration. We hence need to fulfill $\lambda \gamma = \lambda'$ for some 
$\lambda'$ in the set of outliers. Accordingly, the solutions are given by 
\cref{eq:gamma_j} in the main text.

The case $\lambda = \lambda'$ and hence $\gamma=1$ was omitted in the discussion above. 
However, one observes from \cref{eq:gamma_step} after insertion of
$\lambda = 1 / \langle\phi'\rangle$ that $\gamma = 1$ would necessitate 
$\mathbf{n}^T\!\left(\mathds{1} - J / \lambda\right)^{-2}\mathbf{m}=0$. 
According to \cref{eq:eigval_eq}, the eigenvalues $\lambda$ are the 
roots of the function 
$f(\lambda) = \mathbf{n}^T\!\left(\mathds{1}\lambda - J \right)^{-1}\mathbf{m} \,-\, 1$. 
The function has the derivative 
$\mathrm{d}f/\mathrm{d}\lambda = -\mathbf{n}^T\!\left(\mathds{1}\lambda - J \right)^{-2}\mathbf{m}$, 
which only vanishes if the roots have multiplicity larger than 1.
Thus $\gamma = 1$ is only a solution if the corresponding 
$\lambda$ has algebraic multiplicity larger than 1.

\section{Stability for rank-two perturbation}%
\label{sub:stability_for_rank_2_perturbation}
The stability of a fixed point in the case of a rank-two perturbation 
can be evaluated similarly to the rank-one case. One simply applies
the matrix determinant lemma once more on the stability matrix.
Let $\lambda = 1 / \langle\phi'\rangle$ be the outlier corresponding to the 
fixed point under consideration. All mean field quantities will hence 
implicitly depend on $\lambda$, even though we omit this dependency in 
the notation. 
The quadratic equation for the stability eigenvalues $\gamma$ reads
\begin{equation}
    \label{eq:stab_eigvals_r2}
    0 = 
    \gamma^2 - \gamma \mathrm{Tr} \tilde{Q}_\gamma + \det(\tilde{Q}_\gamma) \,,
\end{equation}
with the mean field form of the stability matrix 
\begin{equation}
    \tilde{Q}_\gamma = 
    \begin{bmatrix}
        \mathbf{n}^T\!\tilde{M}_\gamma \mathbf{m} &
        \mathbf{n}^T\!\tilde{M}_\gamma \mathbf{u} \\
        \mathbf{v}^T\!\tilde{M}_\gamma \mathbf{m} &
        \mathbf{v}^T\!\tilde{M}_\gamma \mathbf{u}
    \end{bmatrix} \,,
\end{equation}
and $\tilde{M}_\gamma = R'(\mathds{1} - J R'/\gamma)^{-1}$.
As above, \cref{eq:gamma_step}, we have
\begin{equation}
\begin{split}
    \mathbf{a}^T\!\tilde{M}_\gamma \mathbf{b} &=
    \mathbf{a}^T\!
    \left(\mathds{1} - \frac{\langle\phi'\rangle}{\gamma} J \right)^{-1}
    \left(
        \langle \phi' \rangle \mathbf{b}
        + 
        \frac{\langle \phi''' \rangle}{1 - \frac{q}{\gamma}}
        \mathbf{b}^T\! \mathbf{x} \,\mathbf{x}
    \right) \\
    &=
    \frac{1}{\lambda} \mathbf{a}^T M_{\lambda\gamma} \mathbf{b}
    \\&\quad
    + \frac{\langle \phi''' \rangle \gamma}{(\gamma - q)(\gamma - 1)}
    \mathbf{b}^T (\kappa_1 \mathbf{m} + \kappa_2 \mathbf{u})
    \\&\qquad \times 
    \mathbf{a}^T (\gamma M_\lambda - M_{\lambda \gamma}) 
    (\kappa_1 \mathbf{m} + \kappa_2 \mathbf{u})
    \,,
\end{split}
\end{equation}
for two vectors $\mathbf{a}$ and $\mathbf{b}$
and $M_{\lambda} = (\mathds{1} - J / \lambda)^{-1}$ as 
before, \cref{eq:M_lam}.
The second line is valid in case of an autonomous fixed point, 
c.f. \cref{eq:gamma_auto}.
Evaluating the terms appearing in $\tilde{Q}_\gamma$, we arrive at
\begin{equation}
    \tilde{Q}_\gamma = \frac{1}{\lambda}
    \left[
    Q_{\lambda \gamma} + A \,\Delta Q\, \bm{\kappa}\bm{\kappa}^t 
    \right] \,.
\end{equation}
We abbreviated 
$A = 
\frac{\langle \phi''' \rangle \lambda \gamma }{(\gamma - q)(\gamma - 1)}
$
and 
$\Delta Q = \gamma Q_{\lambda} - Q_{\lambda \gamma}$.
The trace and determinant are then conveniently evaluated as 
\begin{align}
    \det(\tilde{Q}_\gamma) 
    &=
    \frac{\det(Q_{\lambda \gamma})}{\lambda^2}
    \left(
    1 + 
    A \bm{\kappa}^t (Q_{\lambda \gamma})^{-1} 
    \Delta Q \bm{\kappa}
    \right)    
    \,,\\
    \mathrm{Tr}(\tilde{Q}_\gamma)
    &=
    \frac{1}{\lambda}
    \left(\mathrm{Tr}(Q_{\lambda \gamma})
    + A \bm{\kappa}^t \Delta Q \bm{\kappa}
    \right)
    \,.
\end{align}
Inserting these expressions into the stability eigenvalue equation 
\eqref{eq:stab_eigvals_r2} and recalling that $\bm{\kappa}$ is an 
eigenvector of $Q_\lambda$, \cref{eq:kappa_ev}, we finally arrive at
\begin{equation}
    0 = 
    \left[
    (\lambda \gamma)^2 - 
    \lambda \gamma \mathrm{Tr}(Q_{\lambda\gamma}) + \det(Q_{\lambda\gamma})
    \right]
    \left[
    1 - A \bm{\kappa}^t \bm{\kappa}
    \right] \,.
\end{equation}
We hence arrive at the same solutions as in the case of a rank-one
perturbation, \cref{eq:gamma_j,eq:gamma_pm}.

\bibliography{ms}

\end{document}